\documentclass[preprint,review,12pt]{elsarticle}

\usepackage[numbers]{natbib}
\usepackage[hyphens]{url}
\usepackage{enumitem}
\usepackage{hyperref}
\usepackage{rotating}
\usepackage{lipsum}
\usepackage{adjustbox}
\usepackage{pbox}
\usepackage[english]{babel}
\usepackage[section]{placeins}
\usepackage{subcaption}
\usepackage[labelfont=bf,textfont=md]{caption}
\usepackage{dashrule}
\usepackage{tikz,tikz-inet}
\usepackage[binary-units=true]{siunitx}
\usepackage{indentfirst}
\usepackage{array}
\usepackage{wrapfig}
\usepackage{multicol}
\usepackage{multirow}
\usepackage{tabu}
\usepackage{graphicx,url}
\usepackage{microtype}
\usepackage{float}
\usepackage{makecell}
\usepackage{units}
\usepackage{booktabs}
\usepackage{algorithm,algorithmic}
\usepackage{verbatim}
\usepackage{setspace}
\usepackage{siunitx}
\usepackage{amssymb}
\usepackage{amsmath} 
\usepackage{numprint}
\usepackage{colortbl}
\usepackage{longtable}

\usetikzlibrary{graphs,graphs.standard}
\usetikzlibrary{arrows, automata}
\usetikzlibrary{positioning}

\usetikzlibrary{calc,positioning}
\usetikzlibrary{trees,arrows}
\usetikzlibrary{arrows.meta}
\usetikzlibrary{positioning}
\usetikzlibrary{matrix,arrows,decorations.pathmorphing}
\usetikzlibrary{hobby}
\graphicspath{ {imgs/} }

\newcommand\tfig{0.45}
\newcommand\tfigsct{0.30}

\begin{document}

\title{Temporal complex networks modeling applied to vehicular ad-hoc networks}

\author[UNICAMP]{Fillipe Santos\corref{cor}}
\ead{fillipesantos00@gmail.com}
\author[UFAL]{Andre L. L. Aquino}
\ead{alla@laccan.ufal.br}
\author[UNICAMP]{Edmundo R. M. Madeira}
\ead{edmundo@ic.unicamp.br}
\author[UFAL-Ara]{Raquel S. Cabral}
\ead{raquel.cabral@arapiraca.ufal.br}

\cortext[cor]{Corresponding author}

\address[UNICAMP]{University of Campinas.
Campinas, SP, Brazil}
\address[UFAL-Ara]{Federal University of Alagoas - Campus Arapiraca.
Arapiraca, AL, Brazil}
\address[UFAL]{Federal University of Alagoas.
Macei\'{o}, AL, Brazil} 

\begin{abstract}
%%%%%%%%%%%%%%%%%%
%% Alla - new abstract
%%  RSC -- ok ingles
VANETs solutions use aggregated graph representation to model the interaction among the vehicles and different aggregated complex network measures to quantify some topological characteristics. 
This modeling ignores the temporal interactions between the cars, causing loss of information or unrealistic behavior. 
This work proposes the use of both temporal graphs and temporal measures to model VANETs applications. 
To verify the viability of this model, we initially perform a comparative analysis between the temporal and aggregated modeling considering five different real datasets. 
This analysis shows that the aggregated model is inefficient in modeling the temporal aspects of networks. 
After that, we perform a network evaluation through a simulation by considering the impact of temporal modeling applied to the deployment of RSUs. 
First, we compare a solution based on our temporal modeling with a greedy algorithm based on an aggregated model to choose the positions of RSUs. 
In a scenario with 70 RSUs, we have 77\% and 65\% of coverage in the temporal and aggregated model (greedy algorithm), respectively. Second, we evaluate the use of aggregated and temporal measures applied as features in a genetic algorithm. 
The approach with temporal betweenness had the better result with 90\% of the coverage area against 61\% of aggregated one applied to the same scenario.
\end{abstract}


\maketitle

\section{Introduction} \label{introduction} 

%%%%%%%%%%%%%%%%%%
%% Alla - new problem description
A Vehicular ad hoc network (VANET)~\cite{Naboulsi2017} consists of groups of moving or stationary vehicles connected by a wireless network.
According to their communication radius, the communication occurs among the cars or through a fixed infrastructure along the roads. 
In both cases, the connection patterns between their elements have a dynamic behavior with the insertion and removal of links over time.
It is usual to use a graph representation to model the interaction among the vehicles and different complex network measures to quantify some topological characteristics~\citep{Newman:2010:NI:1809753}.
These measures represent a mathematical and computational framework to understand better non-trivial topological features, like the dynamics of growing a network over time, and to enable the characterization, analysis, and modeling of the network topology based on different features like connectivity centrality, cycles, and distances.

Several VANETs applications and network infrastructure solutions use graph modeling and its measures as a static representation called aggregated modeling~\cite{Fiore2008, 9183417, Glacet2015}.
In this modeling, all quick contacts among the vehicles are permanent connections.
For example, consider that we observed three vehicles $v_1$, $v_2$, and $v_3$ on a VANET for ten minutes at one-minute intervals and obtained the following sequence of contacts between them:  
i) $v_1$ with $v_2$ only during minute 1,
ii) $v_1$ with $v_3$ between minutes 2 and 5, and 
iii) $v_2$ with $v_3$ between minutes 6 and 10.
In this case, the aggregated modeling produces a complete graph among the three vehicles, and, consequently, we calculate the aggregated measures directly over this aggregated graph. 
This modeling ignores the temporal interactions between the vehicles, causing loss of information or unrealistic behavior.

%% Alla - checar se as referencias estão corretas
Some applications and infrastructure solutions adopt temporal modeling to represent the graph and preserve the temporal characteristics~\cite{Qiao2017}.
In this case, quick contacts among the vehicles generate connections in a time interval $t$, getting different graphs. 
Considering the previous example with three vehicles, the sequence of contacts in the temporal modeling produces temporal graphs with three disjoint edges $(v_1, v_2)$ at minute 1, $(v_1, v_3)$ during minutes 2 and 5 and $(v_2, v_3)$ during minutes 5 and 10.
However, the solutions calculate the average aggregated measures for all graphs in time interval t, losing global time information about the network.

In this way, the novelty of our proposal is the use of temporal models in conjunction with temporal measures to model and evaluate VANETs.
To the best of our knowledge, we are the only ones that use these measures in VANETs scenarios. 
We use the temporal measures proposed by Kim et al.~\cite{kim2012temporal}: Degree, Betweenness, and Closeness. 
These measures relate node's position and its ability to disseminate information efficiently in the network. 
In addition, they preserve the temporal relationships of the network. 
We perform a comparative analysis between the complete temporal (graph and measures) and aggregated modeling based on these measures.

To quantify the impact of temporal modeling compared with the aggregated one, we consider four static analyses:
i. Quantification of the number of vertices and edges; 
ii. Kolmogorov-Smirnov test (KS-test)~\cite{frank1951};
iii. Hellinger distance~\cite{BASU2010}; and
iv. Visual scatter plot behavior. 
In all cases, we apply the temporal model to follow real scenarios:
\textit{Cologne dataset}~\cite{naboulsi2013} representing the car traffic in an urban area of Cologne, Germany;
\textit{Motorway M40}~\cite{Gramaglia2016} representing part of the intermediate layer of the Madrid city;
\textit{Autovía A6}~\cite{MarieAnge2015} representing the Motorway that connects the city of A Corunã to the city of Madrid;
\textit{Créteil 7am-9am} and \textit{Creteil 5pm-7pm} L\'{e}bre et al.~\cite{MarieAnge2015} representing Créteil, Val-de-Marne (94) in France. 
The results show that the aggregated model is inefficient in modeling the network's temporal aspects.

We also perform a network evaluation through a simulation by considering the deployment of Road Side Units (RSU) application~\cite{moura2018evolutionary}.
In this evaluation, we use both temporal and aggregated models to extract the features used by each algorithm evaluated and compare them using the temporal and aggregated modeling.
We use only the Cologne scenario to perform this evaluation because it is a large-scale dataset that comprises more than 250.000 vehicle routes with varied road traffic conditions.
We first compare our strategy with a greedy algorithm to choose the RSUs' positions. 
This algorithm uses a contact matrix $\mathcal{T}$ to perform the solution. Thus, we generate $\mathcal{T}$ based on aggregated and temporal graphs. 
In a scenario with 70 RSUs, we have 77\% and 65\% of coverage in the temporal and aggregated model, respectively. 
After that, we compare aggregated modeling against the temporal ones as features in the genetic algorithm proposed by Moura et al.~\cite{moura2018evolutionary}. 
This algorithm considers a preprocessing based on different centrality measures. 
Thus, we evaluate both aggregated and temporal measures in specifics scenarios.
The approach with temporal betweenness had the best result with 90\% of the coverage area against 61\% of aggregated one applied to the same scenario.
The evaluation showed that the temporal model is adequate because it truly captures the network behavior.
%%%%%%%%%%%%%%%%%%

We organized the remainder of this paper as follows: 
Section~\ref{sec:realtedWork} gives an overview of the main related work. 
Section~\ref{sec:definitions} describes the temporal VANETs topology modeling. 
Section~\ref{sec:evaluation} shows the evaluations and experiments.
Finally, Section~\ref{sec:conclusion} presents the conclusion and future work.

\section{Related Work}
\label{sec:realtedWork}

%% Alla - reescrevi algumas coisas, especificamente o parágrafo seguinte, e o início de cada um dos parágrafos posteriores
Researchers using a complex network framework in VANETS show the deficiency of availability, connectivity, reliability, and navigability of these networks and reveal the risks of relying on simplistic models for vehicular mobility. 
These studies generally use models that do not consider the vehicular networks' temporal relationships, showing the need to use more realistic models for these networks is necessary. 
Additionally, centrality measures are crucial for understanding the structural properties of complex relational networks and are also relevant for various spatial factors affecting human life and behaviors in cities~\cite{Crucitti2006}. 
Other studies use the measures information to improve the application infer, for instance, using clustering measure in Quality of Service(QoS) or Security decision~\cite{Wahab2013, Wahab2016}.
These theoretical analyses are helpful to understand the structural characteristics of these networks.

% grafos sem métricas de RC 
As mentioned previously, several VANETs applications and network infrastructure solutions use graph modeling and its measures to represent their behavior or extract application features.
In this way, some works model the VANETs as an aggregated graph without considering any complex network measures. 
Elaraby et al.~\cite{Elaraby2021} study the problem of estimating the probability of connectivity in vehicular ad hoc networks (VANETs). 
They discuss the Laplacian eigenvalue and Adjacency exponent approaches to model VANETs connectivity analysis.
Fiore et al.~\cite{8486393} provide helpful guidelines for evaluating and predicting the performance of vehicular networks using an aggregated graph. 
They analyzed a real-world GPS trajectory dataset in a vehicle to vehicle networks.
Feng et al.~\cite{Feng2016} and Chen et al.~\cite{Chen2015} empirically investigate the evolution of spatial-temporal characteristics vehicular environments using mobility data from Shanghai.

%% RSC - Modelo agregado com metricas agregadas. 
We also find additional works using combined aggregated graphs and aggregated complex network measures to model and evaluate VANETs.
Moura et al.~\cite{moura2018evolutionary} and Wang et al.~\cite{wang2017centrality} present different strategies to the RSU problem based on centrality measures.
Naboulsi et al.~\cite{Naboulsi2017} present a topology analysis of two networks from Colony in Germany and Zurique in Switzerland. They show that these networks are poorly connected and, with availability and reliability, very weak.
Youji et al.~\cite{Youji2016} use the average degree, clustering coefficient, and global efficiency to analyze the topology and verify that the edge perturbations have an immediate effect on the characteristics of the topology.
Loulloudes et al.~\cite{Loulloudes2015} provide a ``higher-order'' knowledge in the time-evolving dynamics of vehicular networks in large-scale urban environments through complex network analysis.
Fiore et al.~\cite{Fiore2008} present an in-depth analysis of the topological properties of a vehicular network, unveiling the physical reasons behind the peculiar connectivity dynamics generated by some mobility models.
Crucitti et al.~\cite{Crucitti2006} compare samples of various world cities,  presenting a study of the centrality distributions over geographic networks of urban streets with five measures: centrality, degree, closeness, betweenness, straightness, and information.

% RSC -- grafos temporais e métricas agregadas
On the other hand, different VANETs applications analyze the topology characteristics of mobility from a temporal perspective using temporal graph modeling with aggregated measures.
Diniz et al.~\cite{9183417} use three approaches to model the vehicular networks: instantaneous, aggregated, and time-varying analysis to verify which method can achieve the highest reliability and accuracy. 
They explore two real traces from Rome and San Francisco city and discussed different aggregated metrics that can reveal the fundamental properties of the network.
They verify that instantaneous and time-varying models have the same behavior for the aggregated measures evaluated. 
Qu et al.~\cite{9018063} apply complex network theory to capture the dynamics of the VANET and obtain the analytic definitions of the degree distribution (aggregated).
Celes et al.~\cite{8538726} present a comparison of graphs used in modeling mobility showing the strengths and weaknesses of current approaches in the characterization and analysis of vehicular network topology.
Glacet et al.~\cite{Glacet2015} show that storage and transport mechanisms share the information efficiently.
Several studies explore the temporal aspects of VANETs, for instance, vehicle mobility~\cite{Ho2007, Shioda2008}.

%% RSC -- Modelo temporal com metricas temporais 
The only work that uses, discretely, temporal measures in a specific application is the proposed by 
Qiao et al.~\cite{Qiao2017}.
They explore structural features on a taxi dataset in Beijing with 2927 vehicles. 
They use a temporal model based on the shortest time-ordered path to calculate efficiency and closeness measures. 
They conclude that the results can help create protocols and algorithms to achieve reliability and low latency of communications networks.
However, they do not present a complete analysis of the impact of temporal modeling.
Additionally, our work evaluates five different datasets with distinct characteristics (speed, lane, or congestion levels), allowing us to analyze different views of traffic conditions, and applies the temporal measures in a well-established solution presenting its better performance.  

Table~\ref{tab:related_works} present a qualitative comparison of the literature review with our proposal. 
It lists the manuscripts that consider graphs (aggregated and temporal) and metrics of complex networks (aggregated and temporal) in their applications, evaluations, or comparison.
%Filled cells by $\times$ and \textbullet\ are contributions not considered and considered by the authors, respectively.
In Table, we list the real (R) and synthetic (S) scenarios; aggregated (A) and temporal (T) graphs; degree (D), closeness (C) and betweenness (B) aggregated measures; and degree (TD), closeness (TC) and betweenness (TB) temporal measures.

\begin{table}[htb]
\scriptsize
\centering
\caption{Qualitative comparison of presented works with our proposal.} \label{tab:related_works}
\begin{adjustbox}{width=\columnwidth,center}
\renewcommand\arraystretch{1.5}
\begin{tabular}{l|cc|cc|cccccc}
\toprule
\multirow{2}{*}{\textbf{Literature}} & 
\multicolumn{2}{c|}{\textbf{Scen.}} & 
\multicolumn{2}{c|}{\textbf{Graph}} & 
\multicolumn{6}{c}{\textbf{Measures}} \\
& R & S & A & T & D & C & B & TD & TC & TB \\
\midrule
%% Alla - Verificar se esse grupo usa uma base de dados real ou sintética
Elaraby et al.~\cite{Elaraby2021} & \checkmark & -- &\checkmark & -- & -- & -- & --& -- & -- & --\\
Fiore et al.~\cite{8486393} & -- & \checkmark &\checkmark & -- & -- & -- & --& -- & -- & --\\
Feng et al.~\cite{Feng2016} & -- & \checkmark &\checkmark & -- & -- & -- & --& -- & -- & --\\
Chen et al.~\cite{Chen2015} & -- & \checkmark &\checkmark & -- & -- & -- & --& -- & -- & --\\
Moura et al.~\cite{moura2018evolutionary} & \checkmark  & -- & \checkmark & --  & -- &  -- & \checkmark & -- & -- & -- \\ %\hline 2018
Wang et al.~\cite{wang2017centrality} & -- & \checkmark & \checkmark & -- & \checkmark & \checkmark & -- & -- & -- & --\\ %\hline %2017
Naboulsi et al.~\cite{Naboulsi2017} & \checkmark & -- & \checkmark & -- & \checkmark & -- & \checkmark & -- & -- & -- \\ % 2017
Youji et al.~\cite{Youji2016} & -- & \checkmark & \checkmark & -- & \checkmark & -- &  --&  --&  --&  --\\ %\hline %2016
Loulloudes et al.~\cite{Loulloudes2015} & \checkmark & -- & \checkmark & -- & \checkmark & -- & \checkmark & -- & -- & --\\ %\hline % 2015
Fiore et al.~\cite{Fiore2008} & -- & \checkmark & \checkmark & -- & \checkmark & --  &  -- &  -- &  -- &  --\\ %\hline %2008
Crucitti et al.~\cite{Crucitti2006} & -- & \checkmark & \checkmark & -- & \checkmark & \checkmark & \checkmark &  -- &  -- &  --\\ %\hline %2006
Diniz et al.~\cite{9183417} & \checkmark & -- & \checkmark & \checkmark & \checkmark & \checkmark & \checkmark & -- & -- & -- \\ %\hline %2020
Qu et al.~\cite{9018063} & \checkmark & -- & -- & \checkmark & \checkmark & -- & -- & -- & -- & --  \\ %\hline %2020
Celes et al.~\cite{8538726} & \checkmark & -- & \checkmark & \checkmark & \checkmark & \checkmark & -- & -- & -- & -- \\ %\hline %2018
Glacet et al.~\cite{Glacet2015} & -- & \checkmark & -- & \checkmark & \checkmark & -- &  --&  --&  --&  --\\ %\hline %2015
Ho et al.~\cite{Ho2007} & -- & \checkmark & --& \checkmark & \checkmark & -- &  -- &  -- &  -- &  -- \\ %\hline %2007
Shioda et al.~\cite{Shioda2008} & -- & \checkmark & -- & \checkmark & \checkmark & -- &  -- &  -- &  -- &  -- \\ %\hline %2008
Qiao et al.~\cite{Qiao2017} & \checkmark & -- & -- & \checkmark & --  & -- & -- & -- & \checkmark & --\\ %\hline %2017
\textbf{Present work} & \checkmark & \checkmark & \checkmark & \checkmark & \checkmark & \checkmark & \checkmark & \checkmark & \checkmark & \checkmark \\ %\hline %
\bottomrule
\end{tabular}
\end{adjustbox}
\end{table}

%Additionally, we also find other applications modeled with temporal graphs or measures. The following manuscripts did not enter the Table because it is not related to VANETs.
%Liao et al.~\cite{liao2017ranking} studied ranking algorithms, both static and time-aware, and their applications to evolving networks. They evaluated the impact of network evolution on well-established static algorithms and the benefits of including the temporal dimension for tasks such as network traffic prediction.
%Holme~\cite{holme2015modern} studied how to analyze, model temporal networks and processes to spread infectious diseases and information packages on computer networks.
%Zhou et al.~\cite{zhou2017predicting} predicted social contact patterns from the temporal perspective and proposed a novel approach to improve data forwarding in opportunistic mobile networks.
%Sizemore et al.~\cite{sizemore2018dynamic}  reviewed a set of tools from applied mathematics that offers measures to characterize temporal graphs. 
%However, these applications do not consider that the structure and topology of VANETs change quickly. 
%In this work, we used a temporal complex network model and temporal measures to do an in-depth analysis to understand better the impact of modeling the VANETs as a temporal network.

\section{Temporal VANETs modeling} \label{sec:definitions}

\begin{figure}[!htb]
\centering
\includegraphics[width=0.7\textwidth]{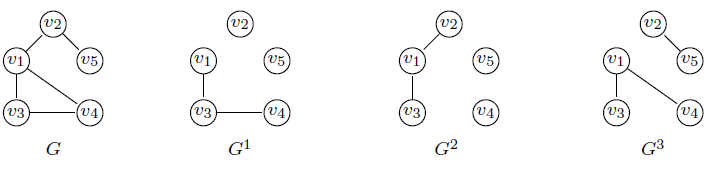}
\caption{Aggregate Graph $G$ obtained from the graphs $G^{1}$, $G^{2}$, and $G^{3}$ of $\mathcal{G}$ with time interval $[1,3]$.}
\label{fig:graph_type}  
\end{figure}

We are modeling the vehicular topology as a temporal graph and characterize it with temporal centrality measures.
The centrality measures are suitable to relate the position of a node in the VANETs with its ability to disseminate information efficiently.
In our model, the vehicular network is a temporal, undirected, and unweighted graph. 
The temporal characteristic defines that the vehicular network is alive from the start time $t_{initial} = 1$ until the end time $t_{final} = T$. 
Let $\mathcal{G} = (V, \mathcal{E})$ be a series of static graphs ${G^1, G^2, \ldots, G^T}$; 
$V=\{v_1, v_2,\ldots, v_M\}$ is the set of vehicles; and $\mathcal{E} = \{E^1, E^2, \ldots, E^T\}$ is the set of disjoint subsets $E^t = \{e^t_1, e^t_2,\ldots,e^t_N\}$. 
Each $E^t$ is the set of edges in $G^t$, with $t \in {[1, T]}$; and $e_a^t = (v_i, v_j)$ is a temporal edge.
The $e_a^t = (v_i, v_j)$ exists, if we have a path between $v_i$ and $v_j$ in the instant $t$, thus edges are added and/or removed over time~\cite{kim2012temporal}. 
The size of $\mathcal{E}$, or the number of temporal edges, is $|\mathcal{E}| = \sum_{i=1}^{T}|E^{i}|$.
Based on this description, an aggregated graph $G = (V,\mathbf{E})$ is the composition $G^1 \cup G^2 \cup \ldots \cup G^T$ of static ones~\cite{kim2012temporal}, i.e., $V=\{v_1, v_2,\ldots, v_M\}$ is the set of vehicles;  and $\mathbf{E} = \{E^1 \cup E^2 \cup \ldots \cup E^T\}$ is the set of aggregated edges.
Figure~\ref{fig:graph_type} illustrates an aggregated graph and the statics used to compose the temporal ($\mathcal{G}$).

Over $\mathcal{G}$, we apply different temporal measures (degree, closeness, and betweenness).
In all definitions of temporal measures, we always consider a generic time interval $[t_x,t_y]$, and we use the aggregated measures as reference~\cite{Costa2008}.

\textbf{Temporal degree centrality} is the number of edges incident to the vertex $v$ in the time interval $[t_x, t_y]$, defined by~\cite{kim2012temporal}
\begin{eqnarray}
\label{eq:temporalDegree}
    \kappa(v)^{[t_x,t_y]} &=& \sum_{i=t_x}^{t_y}\kappa(v)^{i}.
\end{eqnarray}

\begin{figure}[!htb]
\centering
\includegraphics[width=0.7\textwidth]{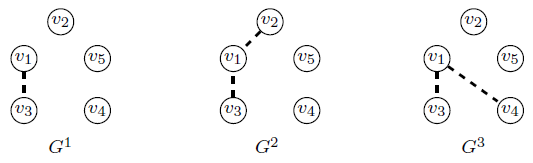}
\caption{Temporal edges to calculate the degree of vertex $v_1$ in temporal graphs in the interval time $[1,3]$.}
    \label{fig:degreeGraph} 
\end{figure}

Figure~\ref{fig:degreeGraph} shows the edges considered to calculate the temporal degree the centrality of vertex $v_1$ in temporal graph $\mathcal{G}$ in the time interval $[1,3]$, in this case $\kappa(A)^{[1,3]} = 5$.

\textbf{Temporal closeness centrality} is a way of detecting nodes that can spread information efficiently through a graph and quantify its average distance to all other nodes. Nodes with a high closeness score have the shortest distances to all other nodes. Kim et al.~\cite{kim2012temporal} defined as: 
\begin{eqnarray}
\label{eq:temporalCloseness}
    C(v)^{[t_x,t_y]} &=& \sum_{i=t_x}^{t_y} \sum_{u \in V}\frac{1}{l(v,u)^{[i,t_y]}}, 
\end{eqnarray}
where $l(u,v)^{[i,t_y]}$ is a temporal shortest path between $u, v \in V,$ i.e., is a sequence of temporal edges which connect $u$ and $v$ in the time interval $[t_x,t_y]$ such that no shorter path exists.

If there is not temporal path from $v$ to $u$ on a time interval $[t,j]$,  we consider $l(v,u)^{[i,t_y]} = \infty$  and we assume $1/\infty = 0$. The normalization factor is $(|V| - 1) \cdot m$, where $m$ is the number of time interval, so $m = 3$. 

\begin{figure}[!htb]
\centering
\includegraphics[width=0.7\textwidth]{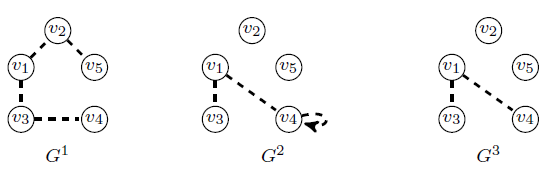}
\caption{Temporal edges to calculate the closeness and betweenness in the interval time $[1,3]$.}\label{fig:closenessGraph} 
\end{figure}

Figure~\ref{fig:closenessGraph} shows the edges considered to calculate closeness of vertex $v_4$. 
To calculate $C(v_4)^{[1,3]}$ we consider the intervals $[1,3],[2,3]$ and $[3,3]$. In the interval $[1,3]$ the minimum paths are $l(v_4,v_1)=2$, $l(v_4,v_2)=3$, $l(v_4,v_3)=1$ and $l(v_4,v_5)=4$, in interval $[2,3]$ node $v_4$ is disconnected so we have a loop at vertex $v_4$ and the minimum paths are $l(v_4,v_1)=2$, $l(v_4,v_2)=\infty$, $l(v_4,v_3)=3$ and $l(v_4,v_5)=\infty$  and in the last interval $[3,3]$ we have  $l(v_4,v_1)=1$, $l(v_4,v_2)=\infty$, $l(v_4,v_3)=2$ and $l(v_4,v_5)=\infty$. In this way $C(v_4) = 0.63$

\textbf{Temporal betweenness centrality} quantifies the extent to which a vertex lies on paths between other vertices. Thus, vertices with high betweenness may have considerable influence over the information passing between nodes. Besides that, removing these nodes from the network can disrupt communications between other vertices~\cite{Newman:2010:NI:1809753}. 
Kim et al.~\cite{kim2012temporal} defined as: 
\begin{eqnarray}
\label{eq:TemporalBetweenness}
    B(v)^{[t_x,t_y]} &=& \sum_{i=t_x}^{t_y}\, \sum_{u,w \in V} \frac{{\sigma(u,w,v)^{[i,t_y]}}}{\sigma(u,v)^{[i,t_y]}},
\end{eqnarray}
where $\sigma(u,v)^{[i,t_y]} > 0$ is the set of temporal shortest paths between $u$ and $v$ in the interval $[i,t_y]$ and $\sigma(u,w,v)^{[i,t_y]}$ is the number of temporal shortest paths that pass through vertex $w$. 
The normalization factor is $(sv\, dv\, m)$, where $m = t_y - t_x$, $sv$ is the number of nodes $s$ that has the shortest path start in $s$ passing through $v$, and $dv$ is the number of nodes $d$ with the shortest path that finishes in $d$ and passes through $v$. 

Figure~\ref{fig:closenessGraph} shows the edges considered to calculate Betweenness of vertex $v_1$. To calculate $B(v_1)^{[1,3]}$ we consider the intervals $[1,3],[2,3]$ and $[3,3]$. In this case, $B(v_1)^{[1,3]} = 7$, $B(v_1)^{[2,3]} = 2$ and $B(v_1)^{[3,3]} = 1$, so $B(v_1) = 0.67$, where the normalization factor to vertex $v_1$ is $36$.  

\section{Evaluations and experiments}
\label{sec:evaluation}

We show the temporal modeling impact by representing the VANETs connectivity with temporal graphs and characterize them with temporal centrality measures.
Initially, we directly apply the temporal modeling in five real data and evaluate it statically.
After that, to perform a dynamic evaluation through a simulation, we consider the deployment of Road Side Units (RSU) application. 

\subsection{Scenarios description}
\label{subsec:scenarios}

In our evaluations, we use five different real datasets.
The datasets have distinct characteristics (speed, lane, or congestion levels), allowing us to analyze different views of traffic conditions. 
Table~\ref{table:general_properties} shows the general properties of each scenario: number of vehicles($|V|$), number of aggregated edges($|\mathbf{E}|$), number of temporal edges($|\mathcal{E}|$).
The datasets represent four different scenarios described as follows.  

\begingroup
\renewcommand\arraystretch{1.5}
\begin{table}[!htb]
\centering
\caption{The general properties of each scenario. }
\scalebox{0.9}{
\begin{tabular}{@{}lccc@{}}
\toprule
\textbf{Scenario}        & $|\textbf{V}|$           &  $|\textbf{E}|$  & $|\mathcal{E}|$ \\ 
\midrule 
\textbf{Cologne}         & \numprint{3558} & \numprint{54268} & \numprint{59769} \\
\textbf{Motorway M4}     & \numprint{2012} & \numprint{33731} & \numprint{93336} \\ 
\textbf{Autov\'ia A6}    & \numprint{1507} & \numprint{24098} & \numprint{68128} \\ 
\textbf{Cr\'eteil 7-9AM} & \numprint{3129} & \numprint{35700} & \numprint{71412} \\  
\textbf{Cr\'eteil 5-7PM} & \numprint{2968} & \numprint{30347} & \numprint{60694} \\ 
\bottomrule
\end{tabular}}
\label{table:general_properties}
\end{table}
\endgroup

\begin{description}
\item[Cologne:] Naboulsi et al.~\cite{naboulsi2013} present a realistic synthetic dataset that reproduces the car traffic in the greater urban area of Cologne, Germany.
The synthetic data closely matches real-world road traffic, both in terms of volume and locality.
They extracted the street layout and features from the OpenStreetMap (OSM) repository\footnote{Available in \url{https://www.openstreetmap.org/}}; the Simulation of Urban Mobility (SUMO)\footnote{ Available in \url{http://sumo.dlr.de/}} software used to simulate the microscopic behavior of individual drivers~\cite{naboulsi2013}.
The dataset spans over 2 hours (6 am-8 am) of a typical workday and encompasses \numprint[km]{4500} of roads in an area of \numprint{400}~km$^2$, with per-second information on the position and speed of vehicles involved in more than \numprint{700000} trips.  
We generated 25 static graphs every 288 seconds (over two hours) for temporal network modeling.

\item[Motorway M40:] 
It is part of the intermediate layer of  Madrid city~\cite{Gramaglia2016, MarieAnge2015}.
It has an average distance of \numprint[km]{10.7} from the city center and traverses both the municipality's peripheral areas and several surrounding minor cities. 
The synthetic traces are composed of one day-long trace describing road traffic over sixteen traces (with  \unit[30]{min}) of vehicular mobility along the road for different days and hour combinations. 
The traces record the position of each vehicle at every \unit[500]{ms}. 
In this scenario, we generated 20 static graphs every 60 seconds for temporal network modeling.

\item[Autov\'ia A6:] 
It is a Motorway that connects the city of La Corun\~a to Madrid. It enters the urban area from the northwest, collecting the traffic demand of the conurbation built along with it~\cite{Gramaglia2016, MarieAnge2015}.
Also, as the M40 trace, the A6 describes road traffic over sixteen traces of \unit[30]{min}, for different days and hour combinations. 
The traces record the position of each vehicle at every \unit[500]{ms} over a \unit[10]{km} road stretch. 
We generated 20 static graphs every 60 seconds for temporal network modeling.

\item [Cr\'eteil 7am-9am and 5pm-7pm:] It based on real traffic of Cr\'eteil in France~\cite{MarieAnge2015}.% (Figure~\ref{fig:france}).
It is a synthetic trace that includes a roundabout with six entrances/exits, two or three lanes roads, one bus road, four changing-lane spots, 15 traffic lights. 
It comprises around \numprint{10000} trips, over rush hour periods of two hours in the morning (7 am to 9 am) and two hours in the evening (5 pm to 7 pm). 
We study each period separately, i.e., each period is a different scenario in our study.
The authors used the OpenStreetMap database to obtain the street layout. 
They derived the traffic demand information on the traffic flow from counting and camera video analysis. 
The traffic assignment of the vehicular flows is performed by Gawron's dynamic user assignment algorithm~\cite{Gawron}, included in the SUMO.
We generated 25 static graphs every 90 seconds for temporal network modeling.
\end{description} 

Each scenario presents different characteristics useful in our analyses:
i. Based on complex network analysis, Cologne presents low connectivity, availability, reliability, and navigability~\cite{naboulsi2013}. 
Also, the structure of this vehicular network has small cliques loosely connected;
ii. Motorway M40 and Autov\'ia A6 have a significant variability of vehicular connectivity over time and space and their invariant correlation with vehicular density. 
The network is more fragmented on A6 than
on M40~\cite{MarieAnge2015}; and 
iii. Cr\'eteil 7am-9am and Cr\'eteil 5pm-7pm present different traffic characteristics in traffic measures (i.e., density, flow, and speed). 

\subsection{Static evaluation of temporal modeling}\label{subsec:experimentalResults}

To quantify the impact of temporal modeling compared with the aggregated one, we consider four static analyses:
i. Quantification of the number of vertices and edges; 
ii. Kolmogorov-Smirnov test (KS-test)~\cite{frank1951};
iii. Hellinger distance~\cite{BASU2010}; and
iv. visual scatter plot behavior. 
In all cases, we model the real scenarios presented above.

In all cases, we use V2V communication among vehicles within a fixed communication radius of \numprint[m]{100}~\cite{basagni2004mobile}.
We perform the static evaluations in \textit{Fedora} 18 (\textit{Spherical Cow}), with 32 processors Intel(R) Xeon(R) CPU $E5-2670$ $\SI{2.60}GHz$, and  $\SI{128}{\giga\byte}$ RAM.
We use the programming languages R~\cite{R} and \textit{Python} (Available on https://www.python.org/) to perform the statistical analysis.
We use the packages \textit{plyr}~\cite{plyr}, \textit{igraph3}~\cite{igraph}, and \textit{time-ordered}~\cite{timeordered}.
They have efficient R/Python modules for manipulation and statistical analysis of graphs and offer methods to incorporate time into network analysis and the construction of temporal graphs.

\begin{figure}[th!]
    \centering
    \begin{subfigure}[]{\tfig\textwidth}
         \includegraphics[width=1\textwidth]{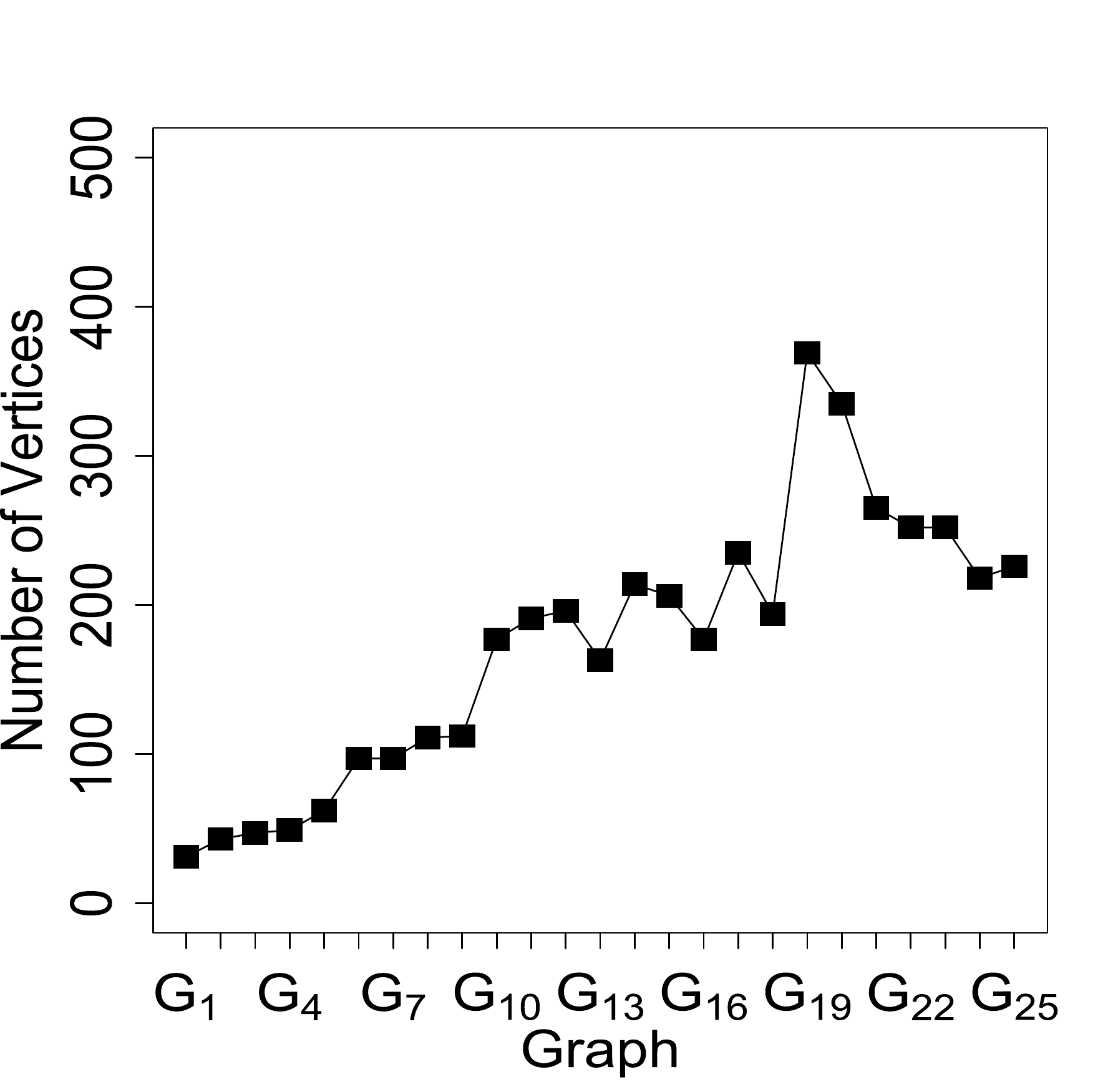}
        \caption{Number of vertices}
        \label{fig:vertex-number-col}
    \end{subfigure}
    \begin{subfigure}[]{\tfig\textwidth}
       \includegraphics[width=1\textwidth]{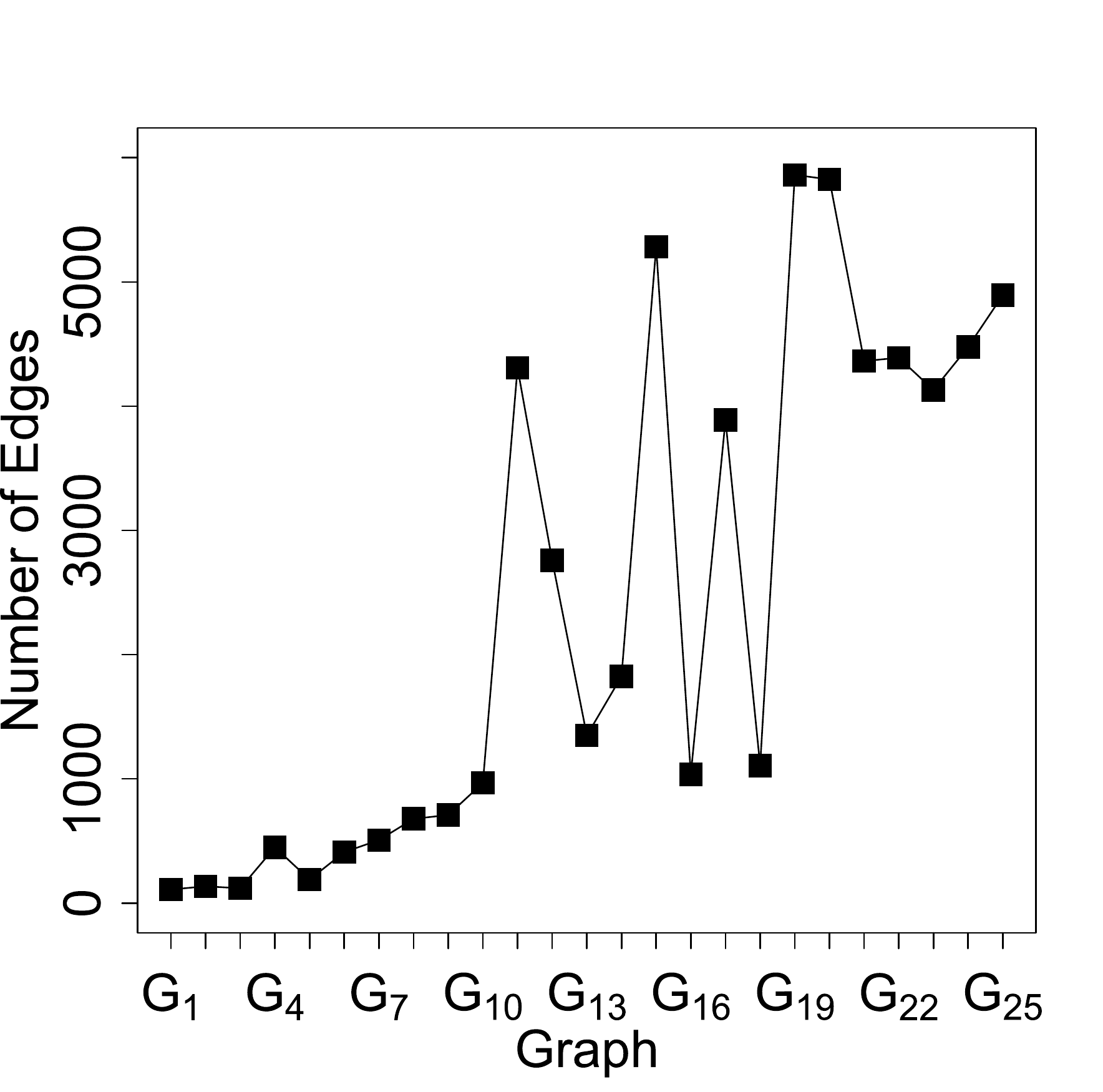}
        \caption{Number of edges}
        \label{fig:edge-number-col}
    \end{subfigure} 
     \caption{Number of static vertices and edges for Cologne scenario in the interval $[0,25]$.}
     \label{fig:number_v_e_cologne}
\end{figure}

In the first analysis, we observe the variations of the network topology.
Figure~\ref{fig:number_v_e_cologne} illustrates the variation of the number of vertices and edges for the Cologne scenario (denser) in a temporal graph in the interval time $[1,25]$.
The aggregated modeling presents a higher number of edges ($\mathbf{E} = \{E^1 \cup E^2 \cup \ldots \cup E^25\}$). 
However, this number has a high variation along time, as observed in the temporal graph in Figure~\ref{fig:edge-number-col}.
This observation indicates that aggregated modeling does not identify topological temporal information.

Based on this initial observation, we perform the second and third analyses.
The KS-test is a non-parametric test to compare two samples, quantifying a distance between two samples' empirical distribution functions.
It verifies if the local measures' distributions ($\kappa$, $C$, and $B$) of the aggregated and temporal modeling are the same. 
The null hypothesis is $$H_0:p_{G}(.) = p_{G^{[0,T]}}(.).$$
In this test, we found a $p$-value $ < \num{2.2e-16}$ for all cases, so we rejected the null hypotheses.
In detail, the Kolmogorov-Smirnov test, $H_0$, the null hypothesis is rejected at the $\alpha$ level if~\cite{frank1951} $\mathcal{D}_{measure}(M) > \delta$, where
\begin{eqnarray}
    \label{eq:ks}
    \delta &=& c(\alpha) \sqrt{\frac{2\times M}{M^2}},
\end{eqnarray}
$M$ is the samples' size of each distribution, i. e., the number of nodes of networks. 
The value of $c(\alpha)$ is $1.36$~\cite{frank1951} representing a $\alpha = 0.05$ (95\% of confidence). 
Table~\ref{table:ks_distances} shows the distance values between aggregated and temporal models for the degree   ($\mathcal{D}_{\kappa}$), closeness ($\mathcal{D}_{C}$), and betweenness ($\mathcal{D}_{B}$) measures; and the value of factor $\delta$ for the five evaluates scenarios. As we can see, $\mathcal{D}_{measure}(M)$ is greater than the factor $\delta$ in all cases. 

\begingroup
\renewcommand\arraystretch{1.5}
\begin{table}[!htb]
%\scriptsize
\centering
\caption{Kolmogorov-Smirnov Distance values between aggregated and temporal graph for the degree ($\mathcal{D}_{\kappa}$), closeness ($\mathcal{D}_{C}$), and betweenness ($\mathcal{D}_{B}$) measures for the five evaluate scenarios.}
\scalebox{0.9}{
\begin{tabular}{@{}lccc|c@{}}
\toprule
\textbf{Scenario} & $\mathcal{D}_{\kappa}$ & $\mathcal{D}_{C}$ & $\mathcal{D}_{B}$ & $\delta$ \\
\midrule
\textbf{Cologne} & 0.9229  & 1.0000 &  0.7745 & 0.0322 \\
\textbf{Motorway M40} &0.9950 & 1.0000 &  0.4602 & 0.0428\\ 
\textbf{Autov\'ia A6} & 0.9827 & 1.0000 & 0.4731  & 0.0495\\ 
\textbf{Cr\'eteil 7am-9am} & 0.9041 & 1.0000 &  0.8590 & 0.0343\\   
\textbf{Cr\'eteil 5pm-7pm} &  0.9154 & 1.0000 &  0.8679 & 0.0353\\ 
\bottomrule
\end{tabular}}
\label{table:ks_distances}
\end{table}
\endgroup

The KS-test statistic is cumulative, i.e., it accentuates low-frequency differences between the measures of aggregated and temporal models.
In the betweenness case, we can distinguish this difference among the presented scenarios. 
For example, the aggregated vs. temporal ($\mathcal{D}_{B}$) representation of the Motorway is more similar than Cologne one.

The other analysis performed, the Hellinger distance, is a distance used to identify the similarity between two probability distributions based on information theory concepts.
This measure quantifies the similarity between two probability distributions. 
Thus it ensures that the distance value is always between 0 and 1 and got a more precise result. 
Consider the discrete random variables $X$ and $Y$ defined on the same sample space $\Omega = \{\xi_1, \xi_2, \dots, \xi_n \}$.
The distributions are characterized by their probability functions $p, q \colon \Omega \to [0,1]$, where $p(\xi_i) = \Pr(X=\xi_i)$ and $q(\xi_i) = \Pr(Y=\xi_i)$.
The Hellinger distance is 
\begin{eqnarray}
 \mathcal{H}^2(p,q) & = & \frac{1}{2}
 \sum_{\xi \in \Omega}\Big( 
 \sqrt{p(\xi)} - \sqrt{q(\xi)}
 \Big)^2.
\label{eq:hellinger}
\end{eqnarray}

By definition, the Hellinger distance is a metric satisfying triangle inequality. The $\sqrt{2}$ in the definition ensure that $\mathcal{H}(p, q) \leq 1$ for all probability distributions. In this work, our sample is the centrality measures probabilities.
To get a probability for each measure, we generate a histogram of proportions with 100 bins for each measure evaluated. We choose the number of bins empirically by testing with 20, 40, 60, 100 (the best fit), 150, and 200 bins.
Table~\ref{table:hellinger} presents the values of Hellinger distance for all scenarios and measures evaluated. 
In general, we observe that the distance captured differences between the models with the measures evaluated.

\begingroup
\renewcommand\arraystretch{1.5}
\begin{table}[!htb]
%\scriptsize
\centering
\caption{Hellinger distance values between aggregated and temporal graph for the degree ($\mathcal{H}^2_{\kappa}$), closeness ($\mathcal{H}^2_{C}$) and betweenness ($\mathcal{H}^2_{B}$) measures for the five evaluates scenarios.}
\scalebox{0.9}{
\begin{tabular}{@{}lccc@{}}
\toprule
\textbf{Scenario} & $\mathcal{H}^2_{\kappa}$ & $\mathcal{H}^2_{C}$ & $\mathcal{H}^2_{B}$ \\
\midrule
\textbf{Cologne} & 0.2524 & 0.4373 & 0.0905   \\
\textbf{Motorway M40} & 0.2533 & 0.4654 &  0.5370 \\ 
\textbf{Autov\'ia A6} & 0.3580 & 0.9603 &  0.4897 \\ 
\textbf{Cr\'eteil 7am-9am} & 0.0077 & 0.7351 & 0.2382 \\   
\textbf{Cr\'eteil 5pm-7pm} & 0.0000  & 0.7407 & 0.3665 \\ 
\bottomrule
\end{tabular}}
\label{table:hellinger}
\end{table}
\endgroup

%% Degree and Hellinger Distance
Values for the degree ($\mathcal{H}^2_{\kappa}$) were smaller than $0.5$ for all scenarios. Results for the Cr\'eteil present lower values ($0.0077$ and $0.0000$) than the previous scenarios. 
This scenario has a high traffic demand and represents a small part of the street. 
Such features affect the network structure and the direct connectivity of vehicles, i.e., high connectivity. 
Based on this measure, the aggregated and temporal models are similar.
%Closeness and Hellinger Distance
Values for the closeness ($\mathcal{H}^2_{C}$) were over $0.7$ for the Autov\'ia A6 and Cr\'eteil 7am-9am and Cr\'eteil 5pm-7pm scenarios. 
The highest value was $0.9603$ in Autov\'ia A6, and this scenario has a small number of vehicles and a significant number of edges distributed in a small area, so the number of edges in the aggregated graph is approximately three times more than a temporal graph.
%Betweenness and Hellinger Distance
The smallest values for the betweenness ($\mathcal{H}^2_{B}$) was $0.0905$ for Cologne. 
This scenario has a very different topology from other ones, where the shortest paths in the temporal graph have a similarity to the shortest paths in the aggregated graph.

It is essential to point out that the measure used to characterize vehicular networks depends not only on the network topology but on the traffic density and network size. Another critical point is the information that we can assess from distributions and their correlation with topological structures. The network's average degree is immediately obtained from the degree distribution, while closeness and betweenness use the distance distribution.
Different from KS-test, we can distinguish better the impact of the aggregated model in the presented scenarios. 

\begin{figure}[!htb]
    \centering
    \begin{subfigure}[]{\tfigsct\textwidth}
        \includegraphics[width=\textwidth]{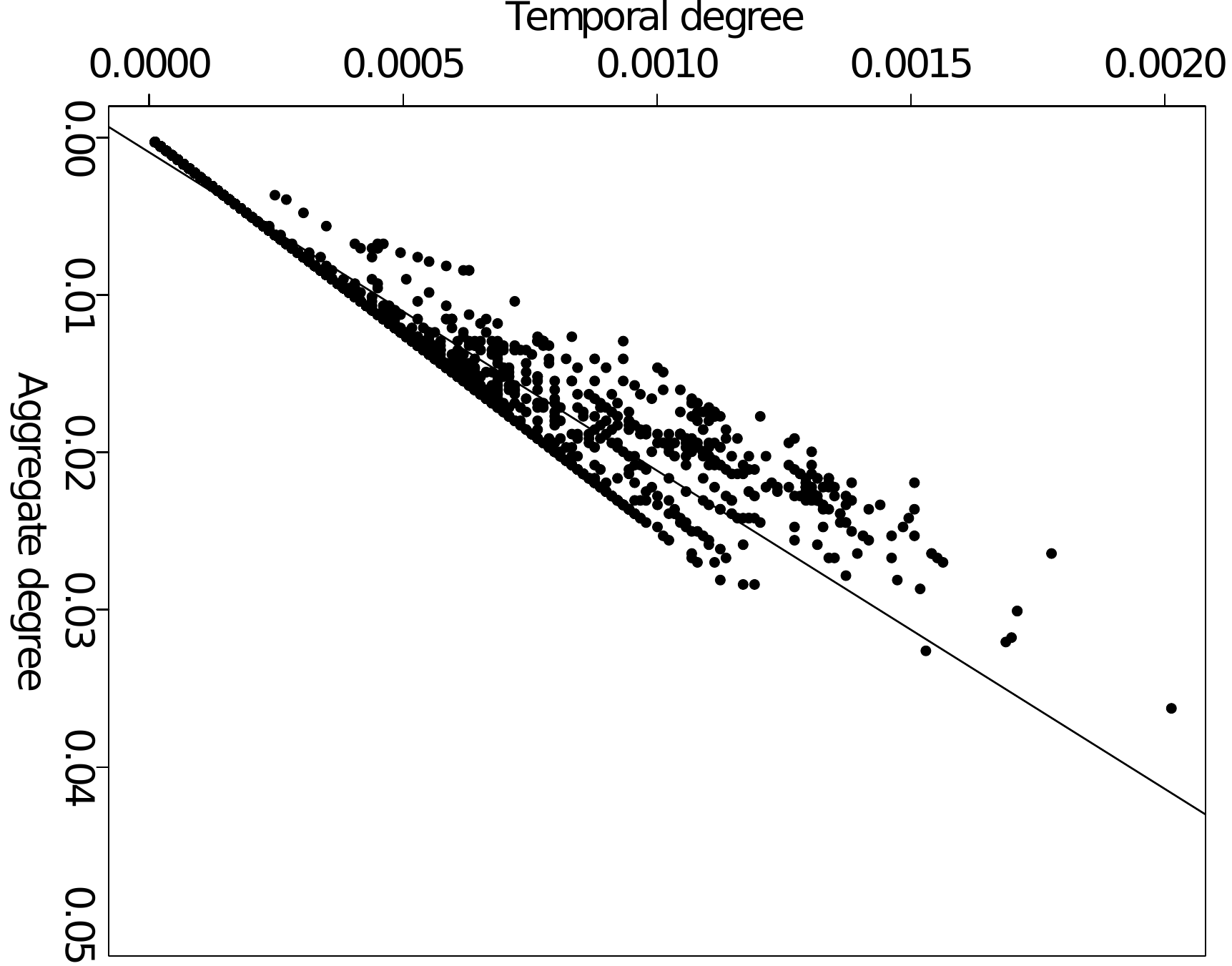}
        \caption{Cologne}
        \label{fig:degree-cologne}
    \end{subfigure} %\hspace{5mm} 
    \begin{subfigure}[]{\tfigsct\textwidth}
        \includegraphics[width=\textwidth]{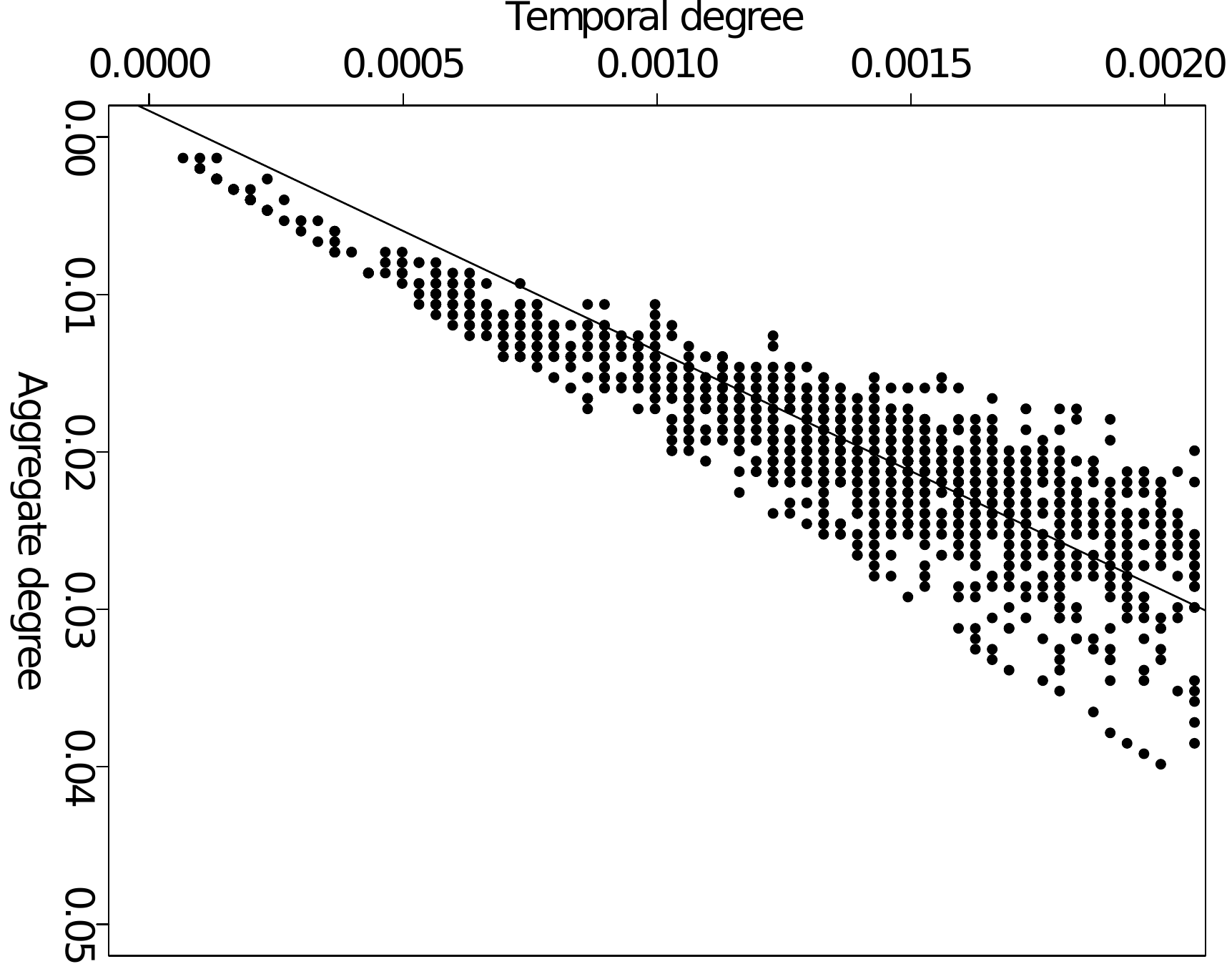}
        \caption{Autovia A6}
        \label{fig:degree-a6}
    \end{subfigure} %\hspace{5mm} 
    \begin{subfigure}[]{\tfigsct\textwidth}
        \includegraphics[width=\textwidth]{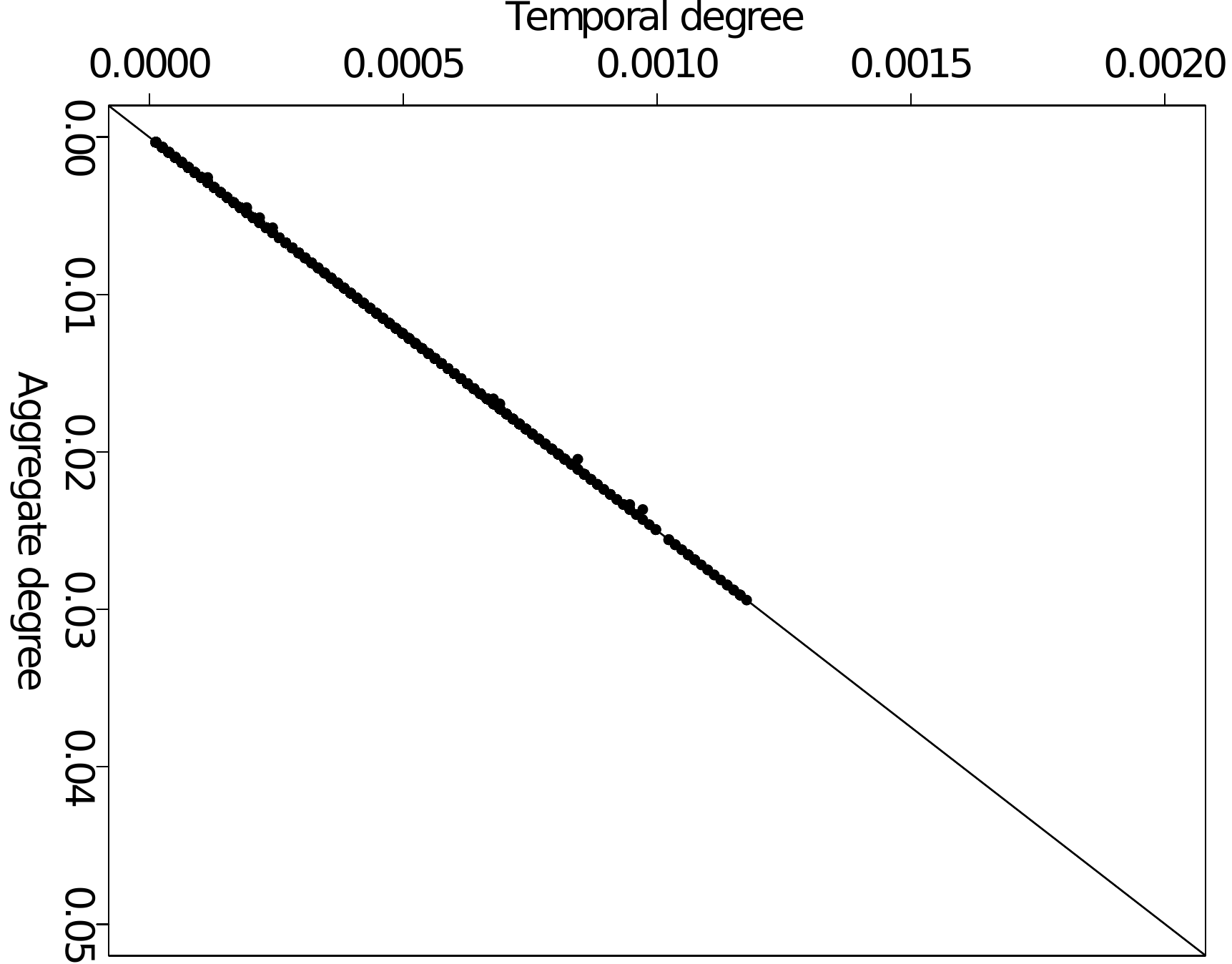}
        \caption{Creteil 7am-9am}
        \label{fig:degree-79}
    \end{subfigure}\\
    \begin{subfigure}[]{\tfigsct\textwidth}
        \includegraphics[width=\textwidth]{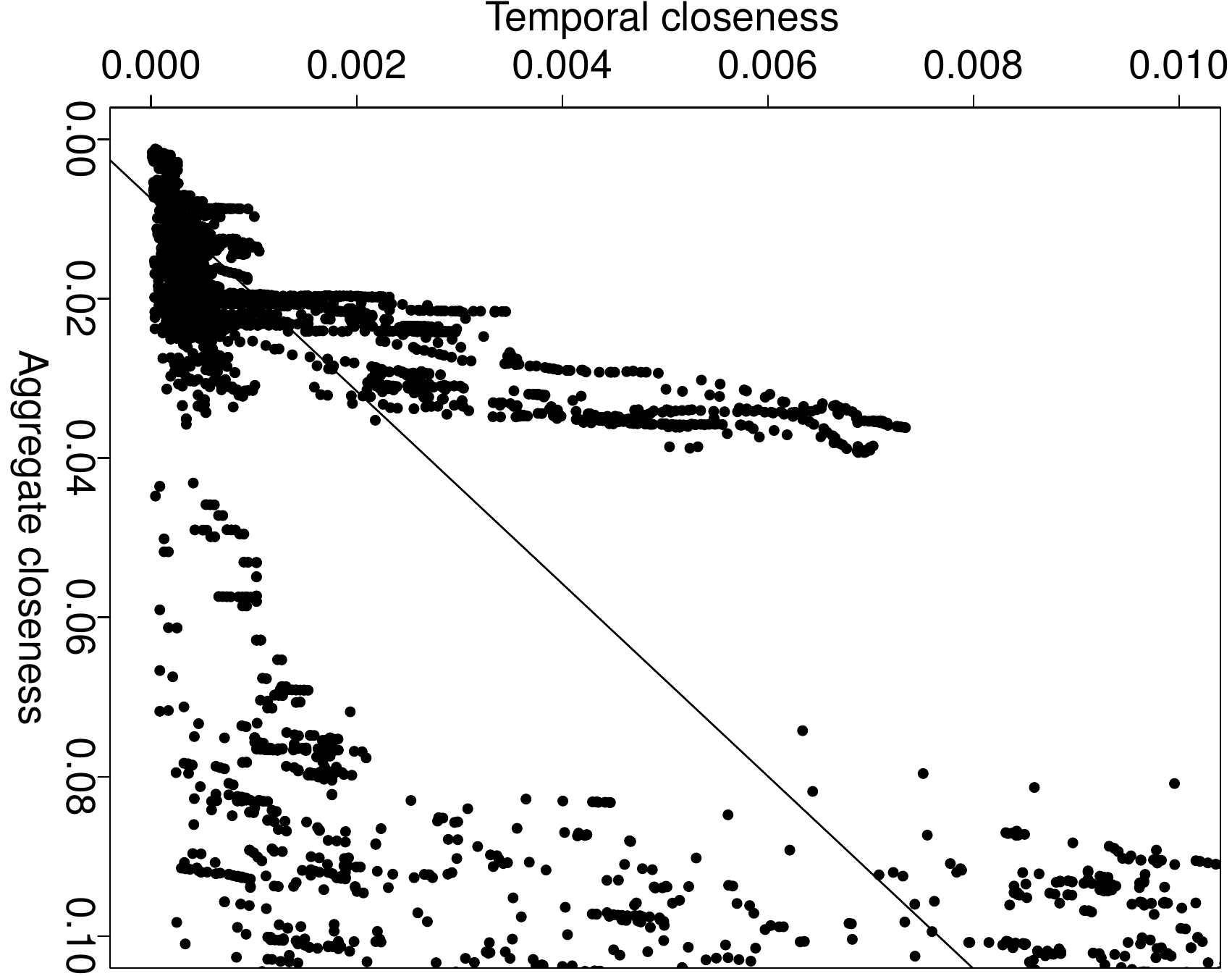}
        \caption{Cologne}
        \label{fig:closeness-cologne}
    \end{subfigure} %\hspace{5mm} 
    \begin{subfigure}[]{\tfigsct\textwidth}
        \includegraphics[width=\textwidth]{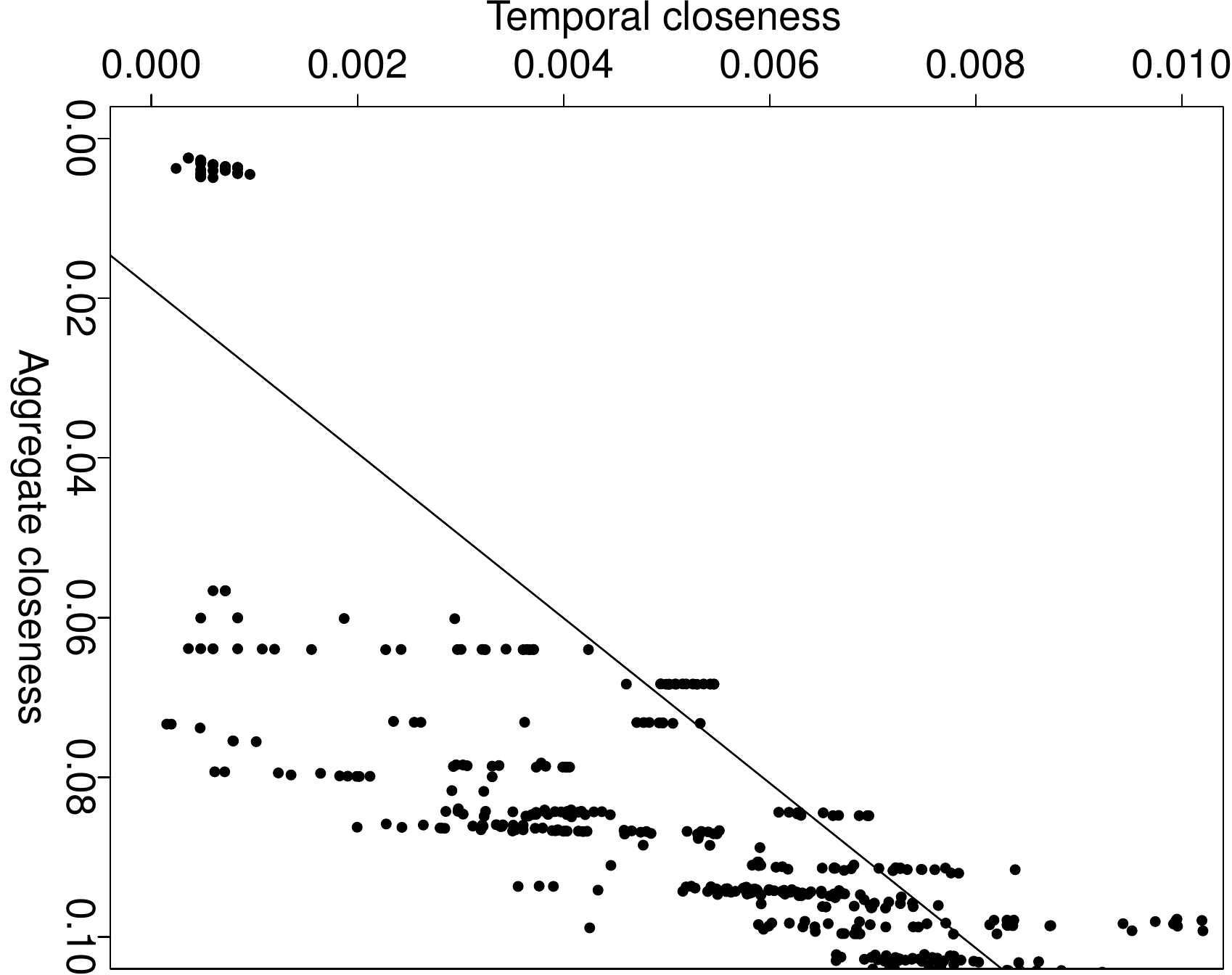}
        \caption{Autovia A6}
        \label{fig:closeness-a6}
    \end{subfigure} %\hspace{5mm} 
    \begin{subfigure}[]{\tfigsct\textwidth}
        \includegraphics[width=\textwidth]{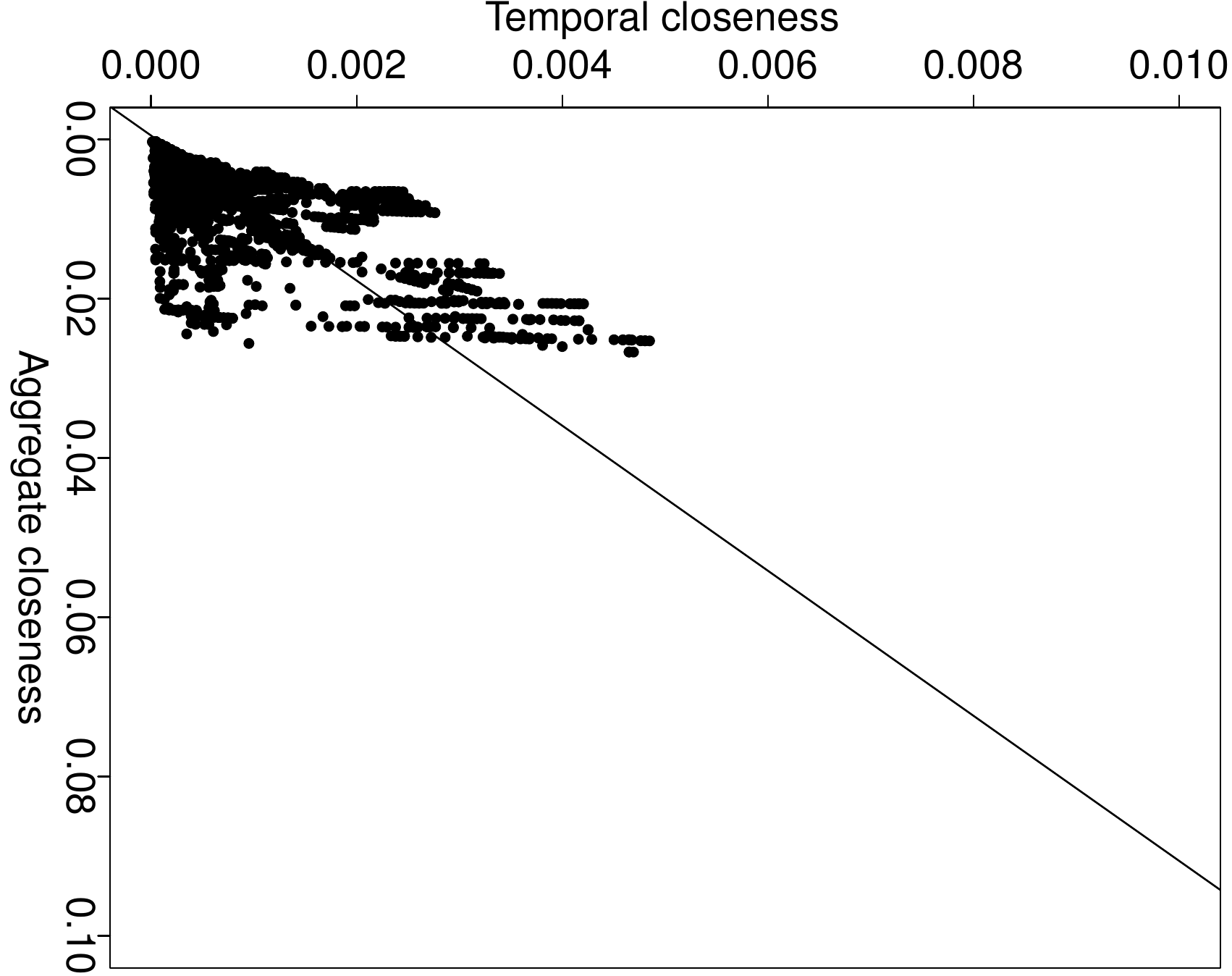}
        \caption{Creteil 7am-9am}
        \label{fig:closeness-79}
    \end{subfigure}\\
    \begin{subfigure}[]{\tfigsct\textwidth}
        \includegraphics[width=\textwidth]{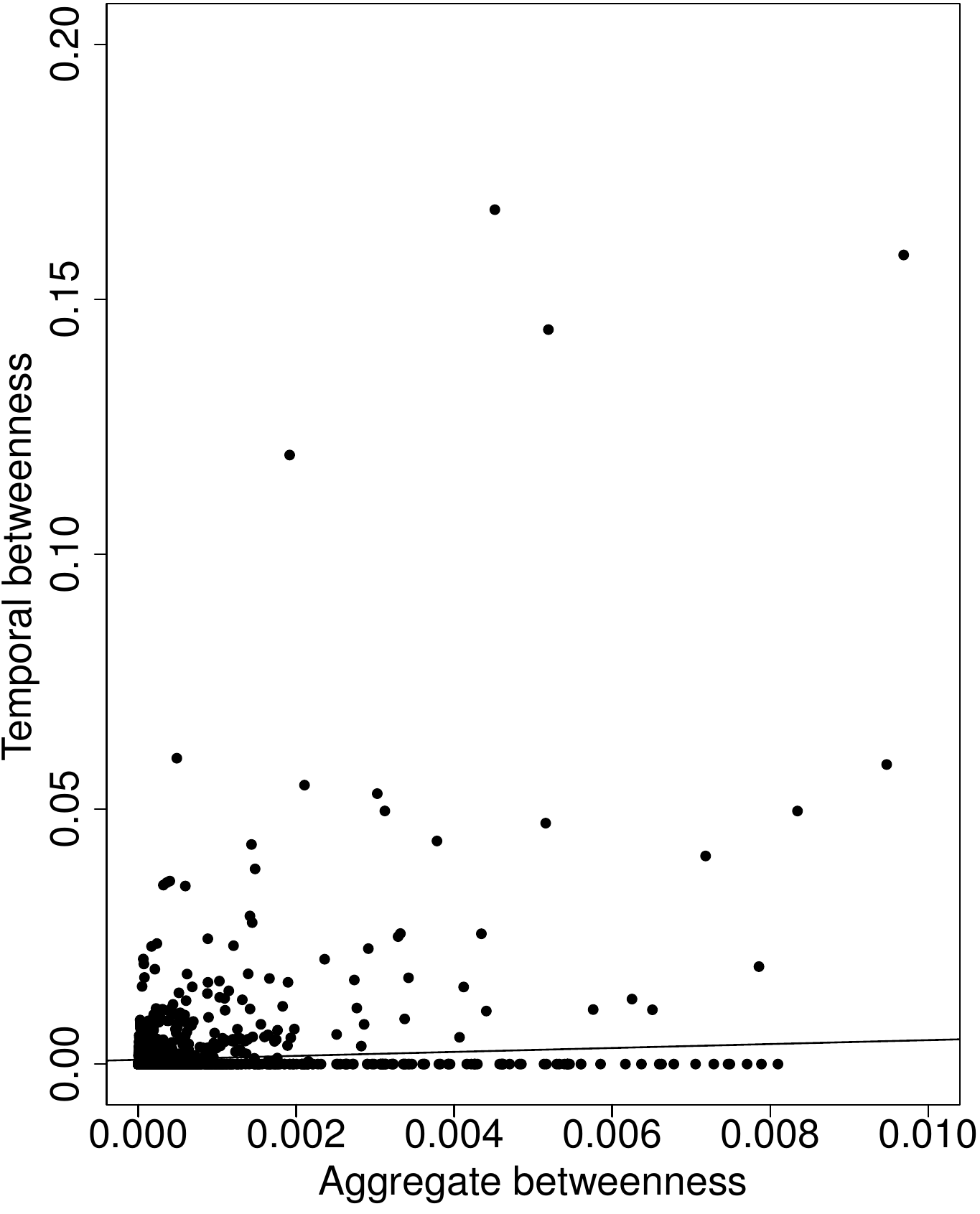}
        \caption{Cologne}
        \label{fig:betweenness-cologne}
    \end{subfigure} %\hspace{5mm} 
    \begin{subfigure}[]{\tfigsct\textwidth}
        \includegraphics[width=\textwidth]{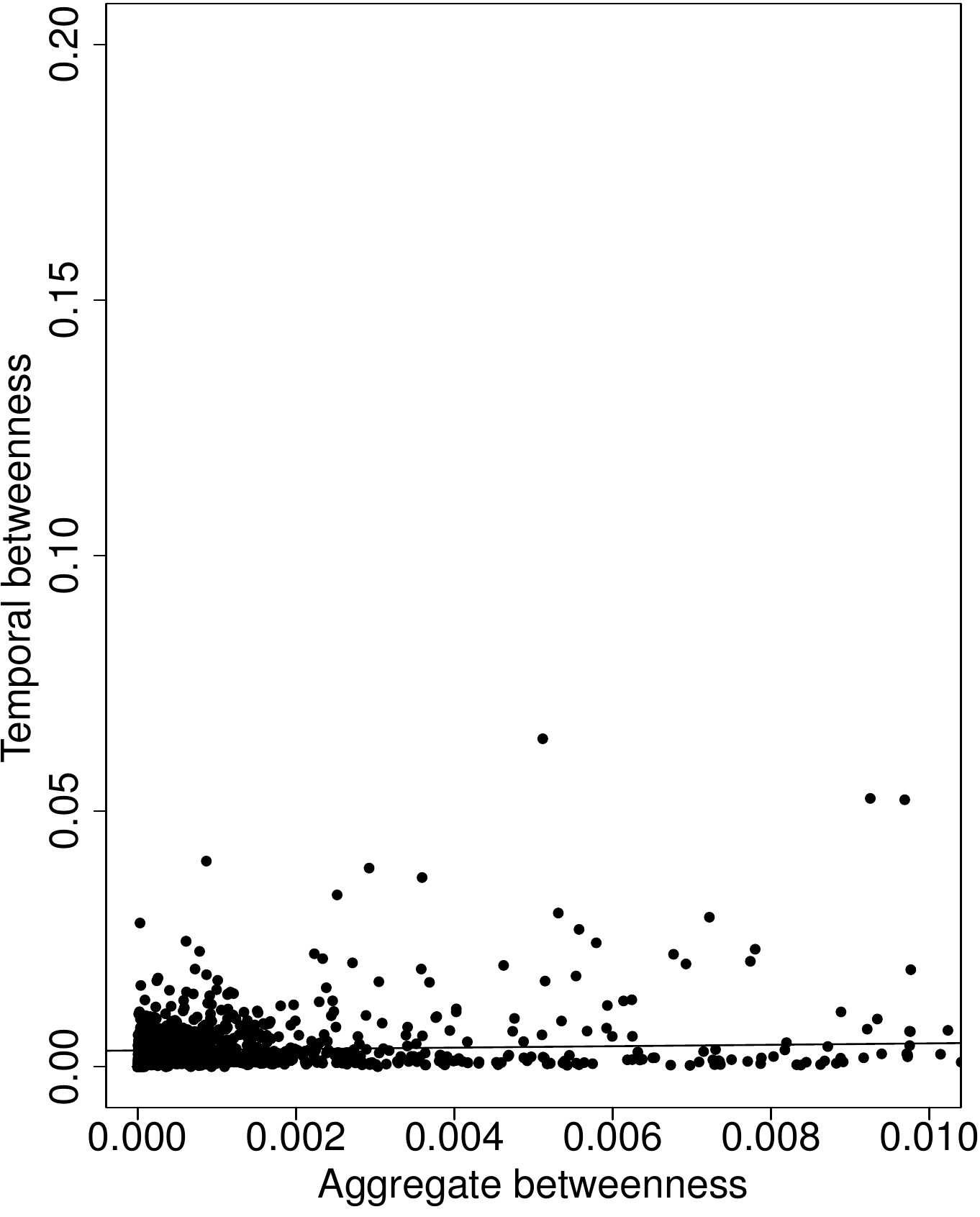}
        \caption{Autovia A6}
        \label{fig:betweenness-a6}
    \end{subfigure} %\hspace{5mm} 
    \begin{subfigure}[]{\tfigsct\textwidth}
        \includegraphics[width=\textwidth]{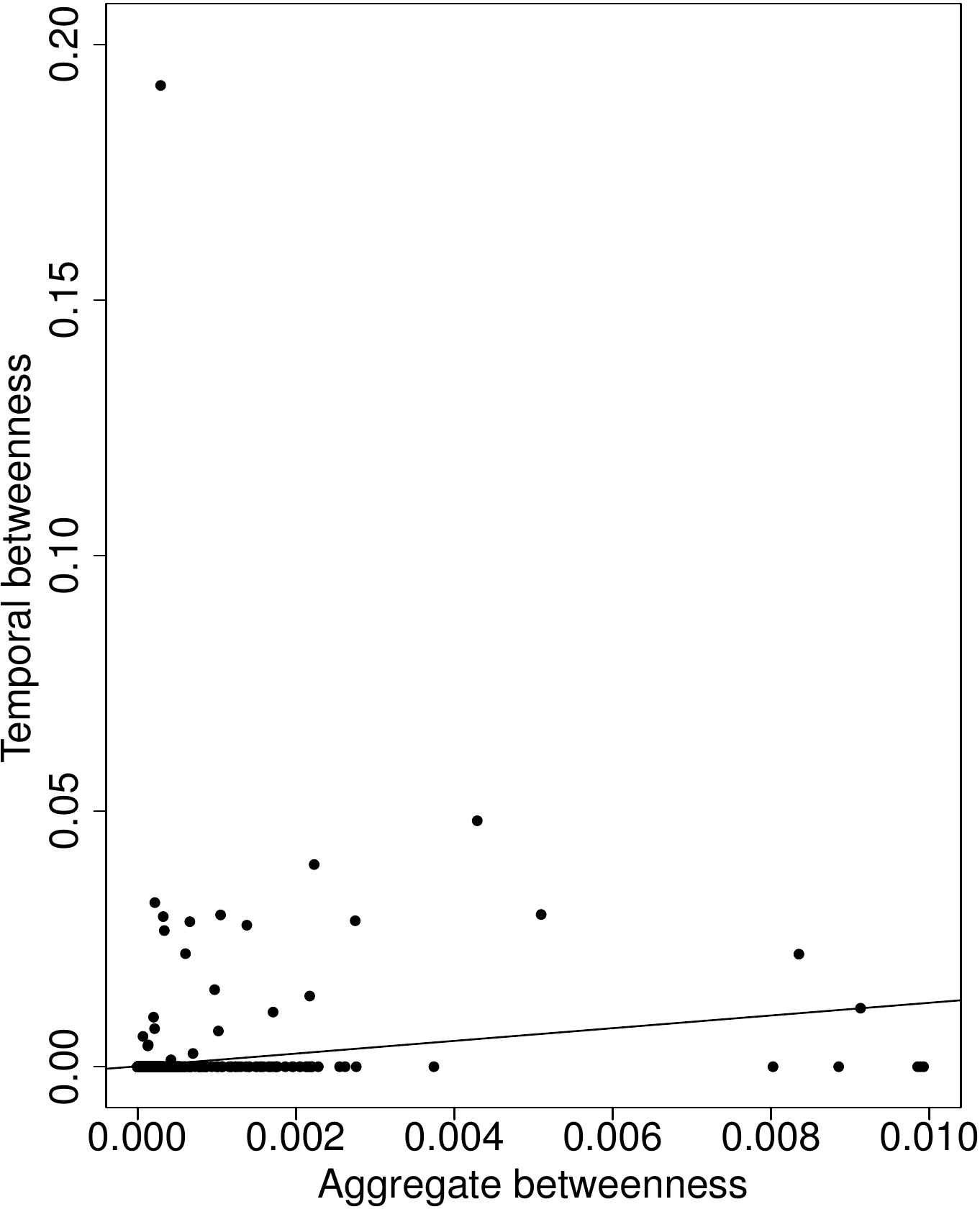}
        \caption{Creteil 7am-9am}
        \label{fig:betweenness-79}
    \end{subfigure}
    \caption{Scatterplot of temporal vs. aggregated models: (a)-(c) Degree; (d)-(f) Closeness; and (g)-(i) Betweenness.}
    \label{fig:scatterplot}
\end{figure}

Finally, we analyze our modeling through a scatter plot.
It shows the measures' correlation between aggregated and temporal modeling. 
We are not interested in whether the correlation between the two measures is positive or negative, but only how strongly the two given measures are numerically related to each other.
Figure~\ref{fig:scatterplot} presents the correlation between the aggregated model (represented on the X-axis) and the temporal one (represented on the Y-axis) regarding degree, closeness, and betweenness, respectively. 

%% Degree scatterplot
As presented in Figures~\ref{fig:degree-cologne}-\ref{fig:degree-79}, there is a positive correlation, but it is not more durable, showing a difference between the aggregated and temporal models regarding degree.
Otherwise, as we can see in Figure~\ref{fig:degree-79}, Cr\'eteil 7am-9am  has a strong positive correlation. 
When the aggregated degree increases, the temporal degree increases as well, so we can show that in this case, the temporal and aggregated networks have the same behavior, stating the result obtained by Hellinger distance. 
It is essential to highlight that the degree is a straightforward measure in studying the topology of networks, and its calculation in both models lost much information (Equation.~\ref{eq:temporalDegree}). 
Some works use degree nodes to determine route strength~\cite{hashim2013new, wang2013passcar, zheng2014reliable}.
In the aggregated graph, a vertex can be identified with a high connection in the wrong way since, by considering the temporal order of the edges, the high link is in a single moment.
Thus, by aggregating the network, temporal relations are ignored, and we consider weak vertices as a great disseminator of information.

%Closeness scatterplot
Figures~\ref{fig:closeness-cologne}-\ref{fig:closeness-79} present the correlation between the aggregated and temporal closeness. 
A vehicle with a high value of temporal closeness centrality indicates that we can quickly reach other vehicles from it.  
Thus, vehicles that have high closeness centrality values can be selected as influence nodes to forward information.
%\cite{qiao2017empirical}.
In this case, we identify no relationship between aggregated and temporal closeness because there is no clear relationship between two quantitative variables; this is because the shortest paths in an aggregated graph are entirely different in the temporal dimension. 
The paths are shortest in the aggregated graph because they abstract the time and interaction between the vertices. 
Also, the paths are shorter because the temporal interaction between the vertices is overlapping. Thus, we can conclude that the shortest paths in a temporal graph follow the links' chronological order, which does not coincide with the shortest paths in an aggregated graph.

%Scatterplot Betweenness
Figures~\ref{fig:betweenness-cologne}-\ref{fig:betweenness-79} illustrate a relationship between the aggregated and temporal betweenness to Cologne, Autov\'ia A6, and Cr\'eteil 7am-9am, respectively.
In this case, the scatter plot's non-structural appearance leads to the summary conclusion that there is no apparent correlation between aggregated and temporal betweenness. 
Nodes with high betweenness have greater control and considerable influence over the network, as more information will pass through these nodes, making them essential for directing network information exchange.
Note, in Figure~\ref{fig:betweenness-cologne},  the behavior of betweenness in the Cologne scenario.  
There is no relationship between variables, but Hellinger's distance is $0.0905$, which contradicts the result seen graphically; this happens because, in this scenario, there are a tiny amount of nonzero values.  
This behavior impacts the speed of the information dissemination process that we overlooked when network aggregation occurs, i.e., a node can always take part in all the shortest paths in the graph.  On the other hand, in the temporal graph, this iteration can occur in a short time.
Thus, the aggregated graph abstracts temporal edges that influence the network's topology and are only visible when modeled as a temporal one.

\subsection{Dynamic evaluation of temporal modeling}
\label{sec:rsu}

We hypothesize that the temporal model is more efficient than the aggregated model. 
In this way, this work considers the deployment of Road Side Units (RSU) application to quantify the impact of temporal modeling compared with the aggregated one in a dynamic and simulated environment. 
A solution to deploy RSUs provides affordable Internet access to connected vehicles and extends the vehicular network coverage (VANETs)~\cite{Naboulsi2017}. 
The deployment of RSUs is essential to design an efficient dissemination system for vehicular environments. 
Several solutions use the vehicular network topology to decide where to install the RSUs. 
The vehicular topology is naturally temporal, i.e., it changes at each instant time.
However, the vehicular topologies are modeling as an aggregated graph. 
We identify previously that this assumption does not allow an adequate representation. 
Thus, we evaluate this application modeling the topology as a temporal graph instead of the usual one.

There are different strategies adopted in the literature to find RSUs positioning in VANETs~\cite{moura2018evolutionary}. 
Generally, they want to choose a set of intersections that optimize the vehicles' coverage time at the dissemination points.
This problem considers an area of an urban road topology, a set of vehicles that transit over the area considered during a given period, a group of intersections between the roads, a subset of vehicles passing each intersection with a minimum contact time of $\tau$. 
The goal is to select $k$ intersections to install a set of RSUs to maximize the number of vehicles covered during a time threshold. 

We apply the classical greedy algorithm to deploy the RSUs problem~\cite{trullols2010planning} considering temporal modeling.
Our goal is to show the greedy algorithm robustness when we model the VANET topology as a temporal graph. 
We use only the Cologne scenario to perform the analysis because it is a large-scale dataset that comprises more than 250.000 vehicle routes with varied road traffic conditions.

\subsubsection{Greedy solution overview}

Many solutions are available in the literature to maximize the number of vehicles covered by $k$ RSUs. 
We use the traditional approach proposed by Trullos et al. \cite{trullols2010planning}. 
The authors model the problem as a Maximum Coverage with Time Threshold Problem (MCTTP).
Let $V = \{v_{1},\ldots, v_{M}\}$ be a set of vehicles that transit over the area considered during a given period; $\Im = \{S_{1},S_{2},\ldots,S_{I}\}$ is a collection of $I$ sets, where $I$ is the number of intersections between the roads; each set $S_i \subset V$ includes all vehicles that cross intersection $i$ at least once over the observation period. 
Let $\mathcal{T}$ be a matrix $I \times M$, where $I$ is the number of intersections, and $M$ is the vehicle number. 
We generate the contact matrix $\mathcal{T}$ in different ways to aggregated ($G$) and temporal ($\mathcal{G}$) graphs. 
\begin{itemize}
    \item We generate the $\mathcal{T}_{G}$ %as depicted by Moura et al.~\cite{moura2018evolutionary}, 
    based on the sum of contact times between vehicles and intersection over a set of aggregated graphs. We aggregated a set of static graphs every 320 seconds and calculated the sum of contact times considering V2I communication.
    \item We generate the $\mathcal{T}_{\mathcal{G}}$ considering the entire time interval (\numprint[s]{7200}), so each element $\mathcal{T}_{\mathcal{G}}(i,j) \geq 0$ is the sum of all contact times that the vehicle $j$ remained at the intersection $i$, i.e., 
    
    $$\mathcal{T}^{[t_x,t_y]}_{\mathcal{G}}(i,j) = \sum_{t_{x} \leq i \leq t_{y}}{\tau^i_{ij}}.$$ 
\end{itemize}

\begin{figure}[!htb]
\centering
\includegraphics[width=0.7\textwidth]{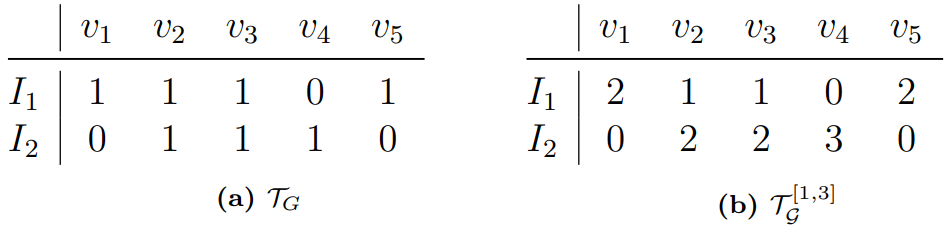}
\caption{Contact matrix to the aggregated and the temporal graph.}
\label{fig:contact_matrix}
\end{figure}

\begin{figure}[!htb]
\centering
\includegraphics[width=0.7\textwidth]{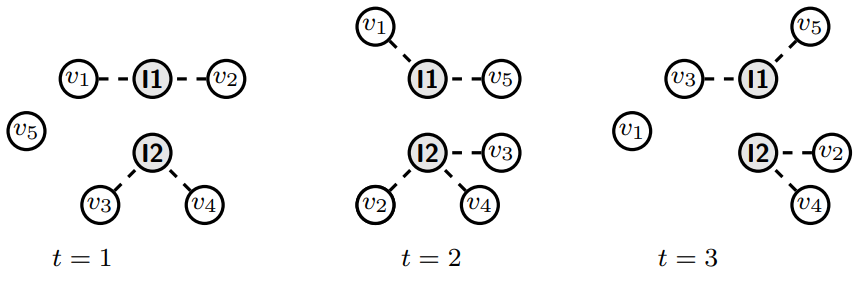}
	\caption{Vehicles and intersections interactions in the time interval $[1, 3]$.}
	\label{fig:graph_contact_matrix}
\end{figure}

Figures~\ref{fig:contact_matrix}-\ref{fig:graph_contact_matrix} show the contact matrices $\mathcal{T}_{\textbf{G}}$ and $\mathcal{T}^{[1,3]}_{\textbf{G}}$ for aggregated and temporal graph, respectively, with $\tau$ = \unit[5]{seconds}. 
In this example, we generate the matrix $\mathcal{T}_{G}$ by aggregating the entire time interval from 1 to 3 seconds.
Otherwise, the matrix $\mathcal{T}^{[1,3]}_{\mathcal{G}}$ is generated consider all contact time in the interval from 1 to 3 seconds. 
As a result, the contact time  to vertex $A$ considering the intersection $I_{1}$ is $\mathcal{T}_{G}(v_1,I_{1}) = 1$ and $\mathcal{T}^{[1,3]}_{\mathcal{G}}(v_1,I_{1}) = 2$.

Trullols et al.~\cite{trullols2010planning} formulate an integer linear problem named Maximum Coverage with Time Threshold Problem (MCTTP). 
The MCTTP selects $k$ intersections to install a subset of RSUs in order to maximize the number of vehicles covered during a time threshold:

\begin{eqnarray}
\label{eq:trullos}
max \sum_{j=1}^{M} \Bigg[ min \Bigg( \tau, \sum_{i=1}^{I}  \mathcal{T}_{ij}\mathcal{Y}_{i} \Bigg) \Bigg]
\end{eqnarray}
s.t.:
\begin{eqnarray}
\label{eq:trullos_rule}
    \sum_{i=1}^{n} \mathcal{Y}_{i} \leq k; \quad \mathcal{Y}_{i} \in \{0,1\} \forall_{i}, 
\end{eqnarray}

Where $\mathcal{Y}_{i}$ is one if there is an RSU in the intersection \textit{i}, and 0 otherwise (Equation~\ref{eq:trullos}). We place an RSU at an intersection to maximize the number of covered vehicles, taking into account a contact time of vehicle up to a maximum value equal to $\tau$ (in this work, we consider $\tau \leq 20s$~\cite{trullols2010planning}). The constraint in Equation~\ref{eq:trullos_rule} instead limits the number of RSUs to $k$. 
One or more RSUs may cover the vehicles;  in the first case, we have only one contact time; in the second, the vehicle contact time is fragmented, i.e., the total contact time is the sum of the contacts among the vehicle and the RSUs.

The greedy algorithm~\ref{alg:guloso}, denoted by MCTTP-g, solves the MCTTP problem. It picks an intersection at each step to maximize the provided coverage time~\cite{trullols2010planning}. The algorithm's input is the number of RSUs $k$ to install, a candidate set $\Im$, the contact time $\tau$, and the matrix of contact time $\mathcal{T}$. The output is the set of RSUs to install $\mathcal{G}$.   

\begin{algorithm}
\caption{MCTTP-g}
\label{alg:guloso}
\begin{algorithmic}[1]
\REQUIRE k, $\Im$, $\tau$, $\mathcal{T}$, 
\ENSURE $\mathcal{G}$ 
\STATE $\mathcal{G} \gets \emptyset$ \label{alg:l1}
\STATE $t_{j} \gets 0, j = \{1,\ldots,M \}$ \label{alg:l2}
\REPEAT
\STATE $W_{i} \gets \sum_{j = 1}^{M} min(\tau-t_{j},  \mathcal{T}_{ij}), i = \{1,\ldots,I\}$ \label{alg:l3}
\STATE Select $S_{i} \in \textbf{S}$ that maximizes $W_{i}$ \label{alg:l4}
\STATE $\Im \gets \Im \setminus S_{i}$ \label{alg:l5}
\STATE $\mathcal{G} \gets \mathcal{G} \cup S_{i}$ \label{alg:l6}
\STATE $t_{j} \gets min(\tau,t_{j} +  \mathcal{T}_{ij}), j = \{1,\ldots,M \}$ \label{alg:l7}
\STATE $k \gets k - 1$ \label{alg:l8}
\UNTIL {$k = 0$ or $\Im = \emptyset$} \label{alg:l9}
\end{algorithmic}
\end{algorithm}

Let $W_{i}$ be the total contact time provided by intersection $i$, considering for each vehicle an amount such that the vehicle's coverage time does not exceed the threshold.
Lines \ref{alg:l4}--\ref{alg:l6} select $S_{i}$ that maximizes $W_i$, removing from $\Im$ and adding a greedy solution $\mathcal{G}$. 
The algorithm ignores the vehicles that have reached the minimum time and complete the transmission in time $\tau$. 
If the vehicle stays time in intersection $i$ not sufficient for transmission, we stored this value in a vector $t_{j}$ (Line~\ref{alg:l7}), and we calculate the time to complete the transmission ($\tau - t_{j}$) in the next iteration (Line~\ref{alg:l3}).

After $k$ intersections selected, or when $\Im = \emptyset$, the algorithm returns the set of selected RSUs $\mathcal{G}$. 
Thus, we consider this algorithm to solve the MCTTP problem and analyze the positioning and coverage of RSUs positioned using the aggregated and temporal graphs. 

\begin{figure}[!htb]
\centering
\includegraphics[width=0.7\textwidth]{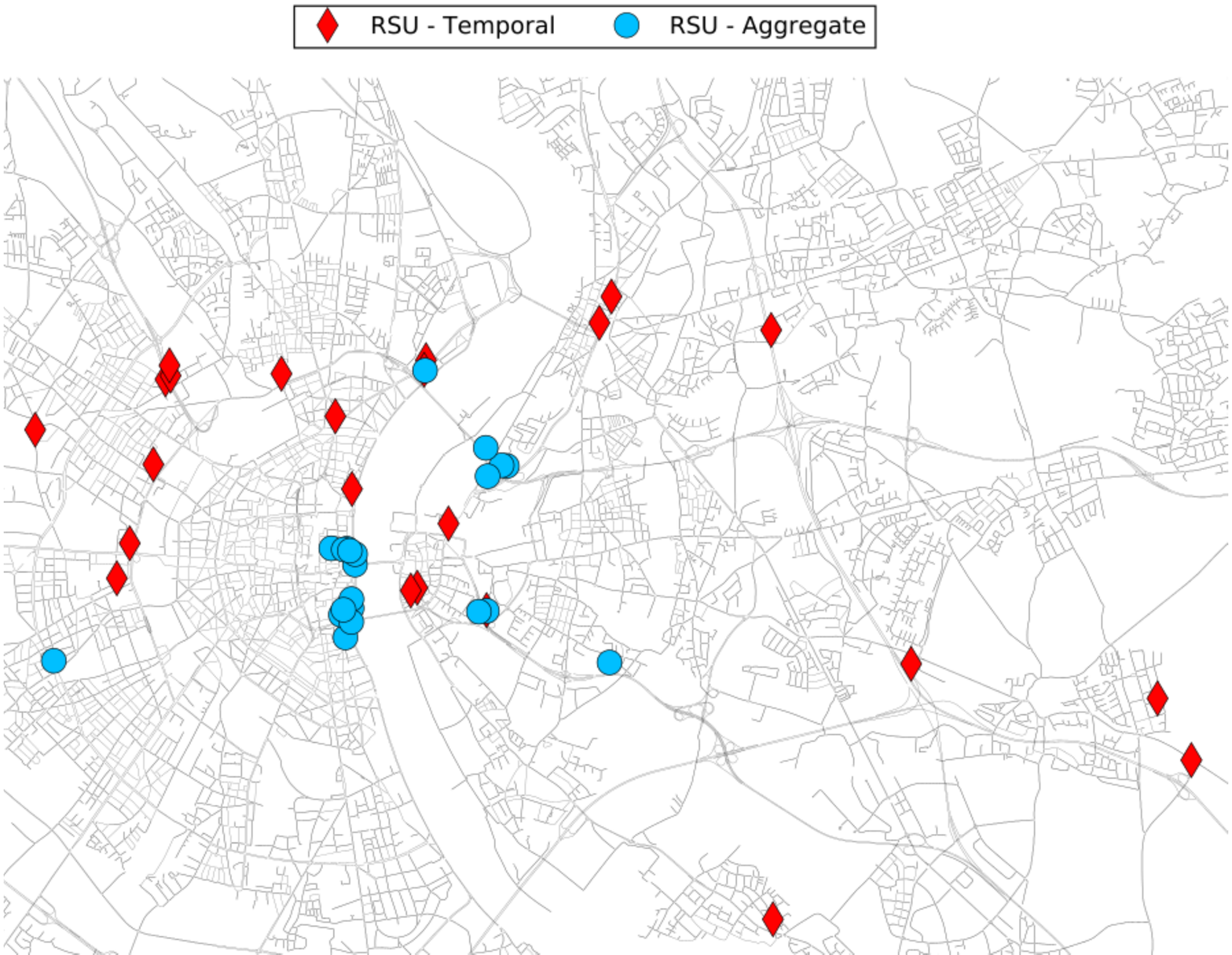}
\caption[RSU Colonia.]{Map view of the central region of Cologne and the installed RSUs to aggregated and temporal models.}
\label{fig:scenario-rsu}
\end{figure}

Figure~\ref{fig:scenario-rsu} shows the map view of Cologne's central region and the installed RSUs for the aggregated model (blue circles) and the temporal model (red diamonds). 
It is the result of the execution of the algorithm~\ref{alg:guloso}.
As we can see, the positioning of the RSUs when we model the scenario with an aggregated graph is concentrated in the central part of the area, covering the same region and the identical vehicles. 
Otherwise, when we model with a temporal graph, the RSUs are more distributed, ensure better coverage. 

\begin{figure}[!htb]
    \centering
    \begin{subfigure}[]{\tfigsct\textwidth}
        \includegraphics[width=1.2\textwidth]{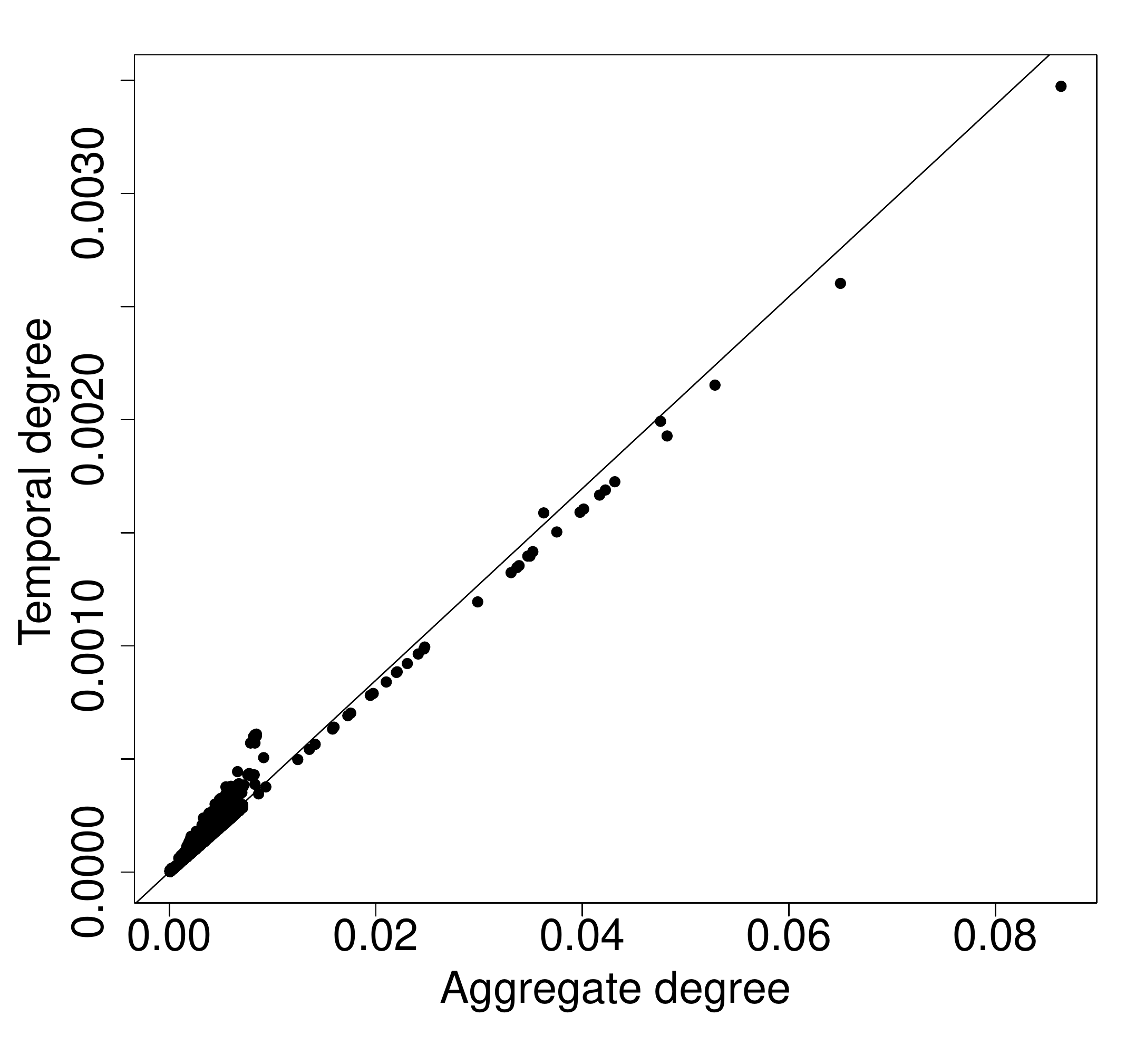}
        \caption{Degree}
        \label{fig:degree-rsu}
    \end{subfigure} \hspace{4mm} 
    \begin{subfigure}[]{\tfigsct\textwidth}
        \includegraphics[width=1.2\textwidth]{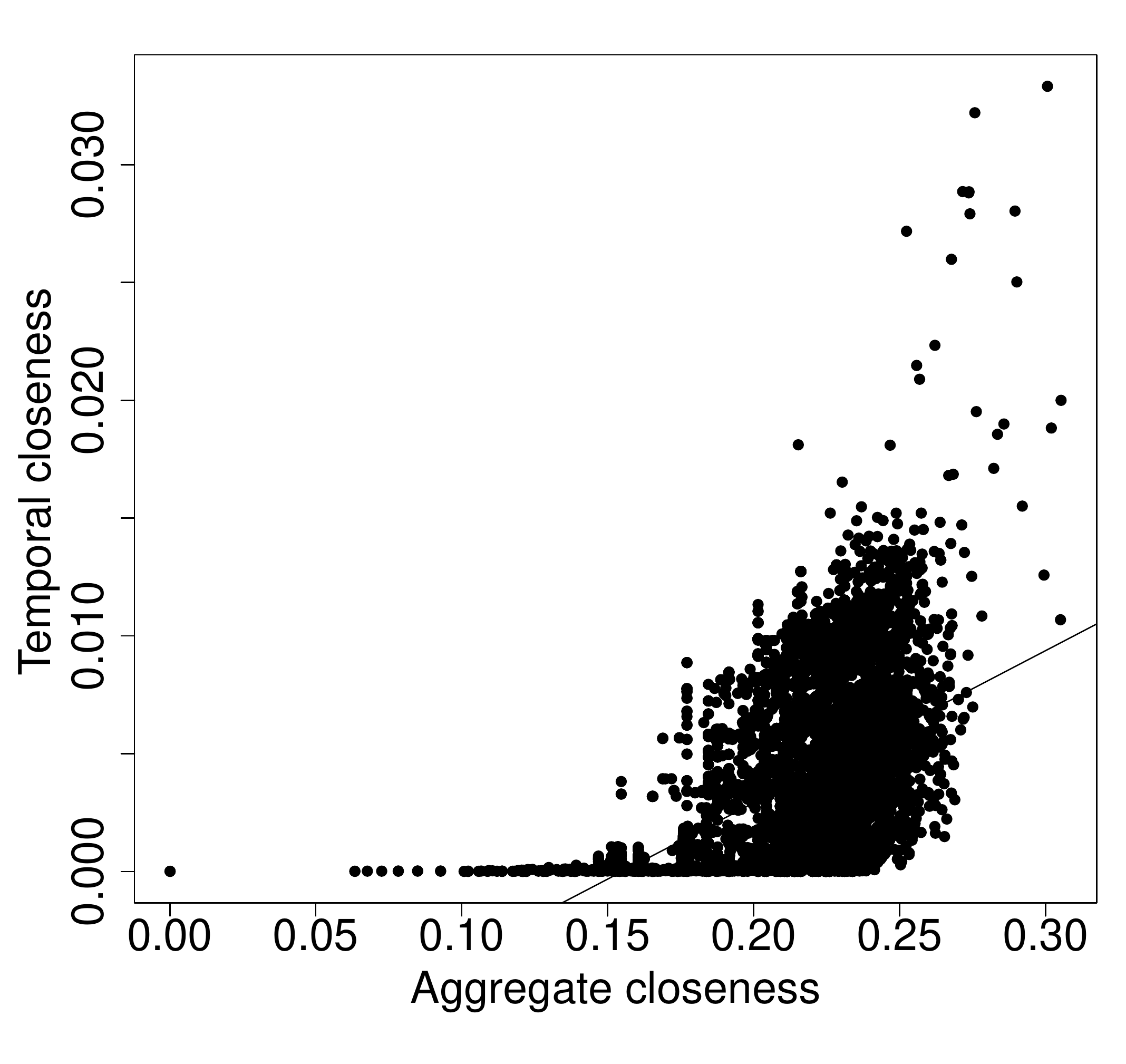}
        \caption{Closeness}
        \label{fig:cls-rsu}
    \end{subfigure} \hspace{4mm} 
    \begin{subfigure}[]{\tfigsct\textwidth}
        \includegraphics[width=1.2\textwidth]{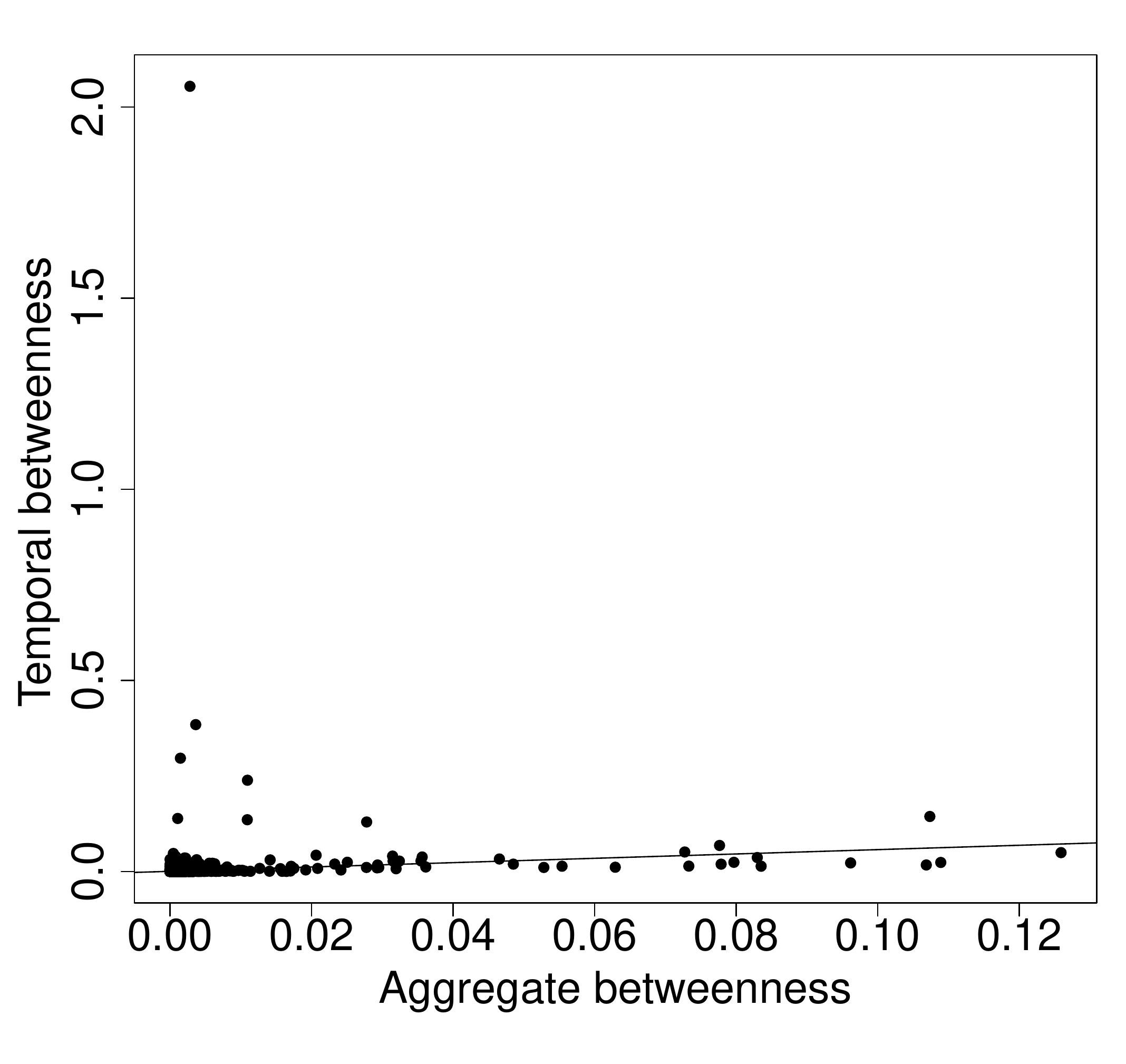}
        \caption{Betweenness}
        \label{fig:btw-rsu}
    \end{subfigure}
    \caption{Scatterplot of temporal vs. aggregated models.}
    \label{fig:scatterplot-rsu}
\end{figure}

Similar to static evaluation, we also use a scatterplot to show the correlation between the aggregated and local temporal measures. 
In this case, we regard $k=500$ RSUs deployed by Algorithm~\ref{alg:guloso} in Cologne scenarios considering the modeling details of both models.
As mentioned before, we are interested in how strongly the temporal and aggregated measures are numerically related.
Figure~\ref{fig:scatterplot-rsu} illustrates the correlation between the aggregated model (represented on the X-axis) and temporal one (represented on the Y-axis) regarding degree, closeness, and betweenness, respectively.

As presented in Figure~\ref{fig:degree-rsu}, there is a positive correlation, but it is not more durable, showing a difference between the aggregated and temporal models regarding degree.
In the aggregated graph, a vertex can be identified with a high connection in the wrong way since, by considering the temporal order of the edges, the high link is in a single moment.
Thus, by aggregating the network, temporal relations are ignored, and we consider weak vertices as a great disseminator of information.

Figure~\ref{fig:cls-rsu} presents the correlation between the aggregated and temporal closeness. 
In this case, we identify no relationship between aggregated and temporal closeness because there is no clear relationship between two quantitative variables; this is because the shortest paths in an aggregated graph are entirely different in the temporal dimension. 
The paths are shortest in the aggregated graph because they abstract the time and interaction between the vertices. 
Also, the paths are shorter because the temporal interaction between the vertices is overlapping. 
Thus, we can conclude that the shortest paths in a temporal graph follow the links' chronological order, which does not coincide with the shortest paths in an aggregated graph.

Figure~\ref{fig:btw-rsu} illustrates a relationship between the aggregated and temporal betweenness.
In this case, the scatter plot's non-structural appearance leads to the summary conclusion that there is no apparent correlation between aggregated and temporal betweenness. 
Nodes with high betweenness have greater control and considerable influence over the network, as more information will pass through these nodes, making them essential for directing network information exchange.
Observe the case for the betweenness in the Cologne scenario. There is an impact on the speed of the information dissemination process that we overlooked when network aggregation occurs, i.e., a node can always take part in all the shortest paths in the graph.  On the other hand, in the temporal graph, this iteration can occur in a short time.
Thus, the aggregated graph abstracts temporal edges that influence the network's topology and are only visible when modeled as a temporal one.

\subsubsection{Network behavior simulation}
This section evaluates the impact of temporal modeling (graph and measures) over the RSUs' deployment.
We conducted the experiments to assess the proposed system using Veins 5.0, an open-source framework for running vehicular network simulations.
Network simulation was performed by OMNeT++ 5.5.1, a discrete event simulator for modeling communication networks, multiprocessors, and other distributed or parallel systems. 
Road traffic simulation was performed by SUMO 1.6.0, which is well-established in traffic engineering, designed to handle large networks. 
The Physical (PHY) and Medium Access Control (MAC) layers are implemented in Veins and based on the IEEE 802.11p (WAVE) standard.
All the simulators are free-source under the Academic Public License.

\begin{figure}[tb]
\space
    \centering
    \begin{subfigure}[]{.47\textwidth}
        \includegraphics[width=\textwidth]{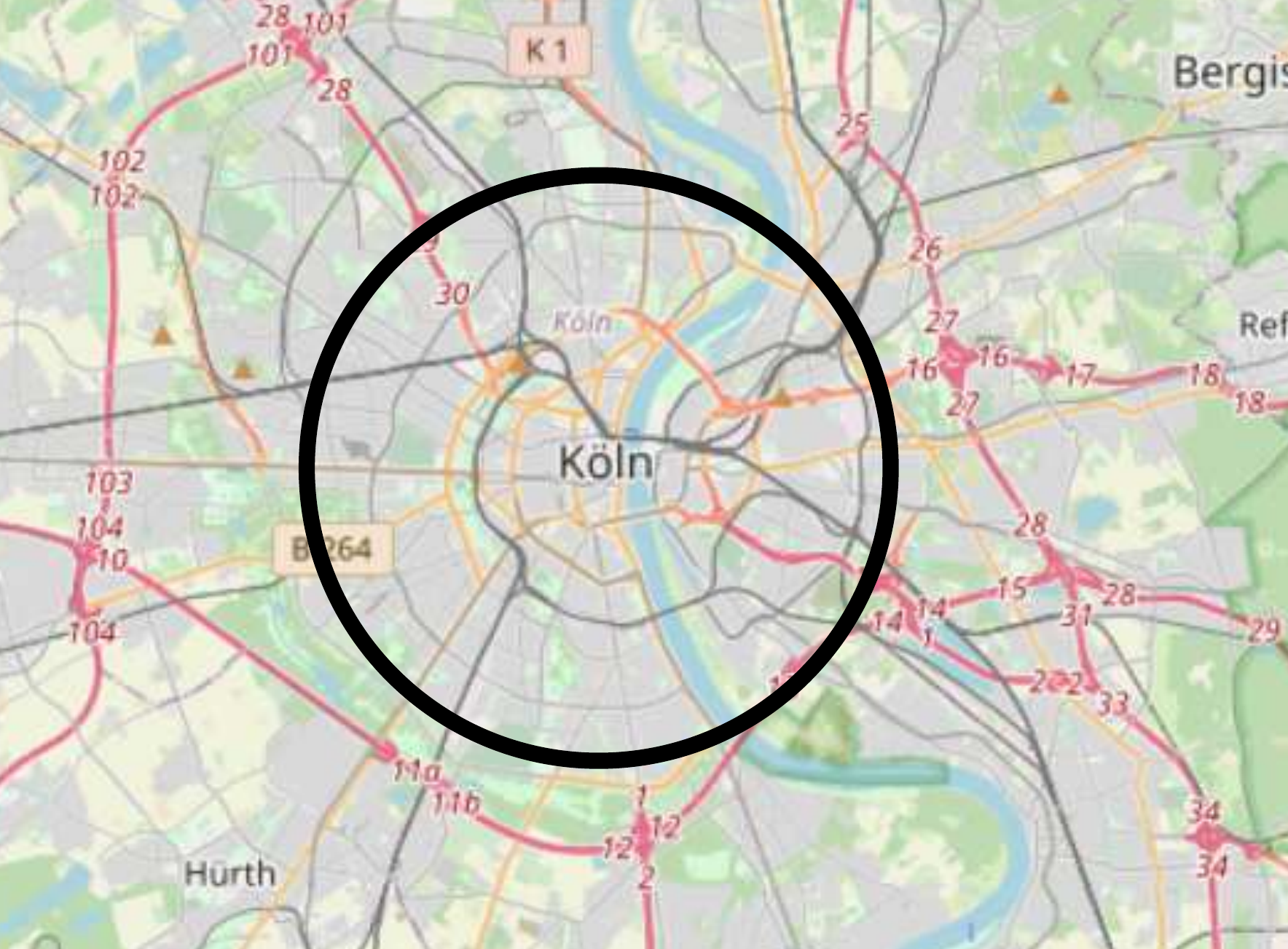}
         \caption{Cologne topology.}
        \label{fig:cologne-topology}
    \end{subfigure}
\space
    \begin{subfigure}[]{.47\textwidth}
        \includegraphics[width=\textwidth]{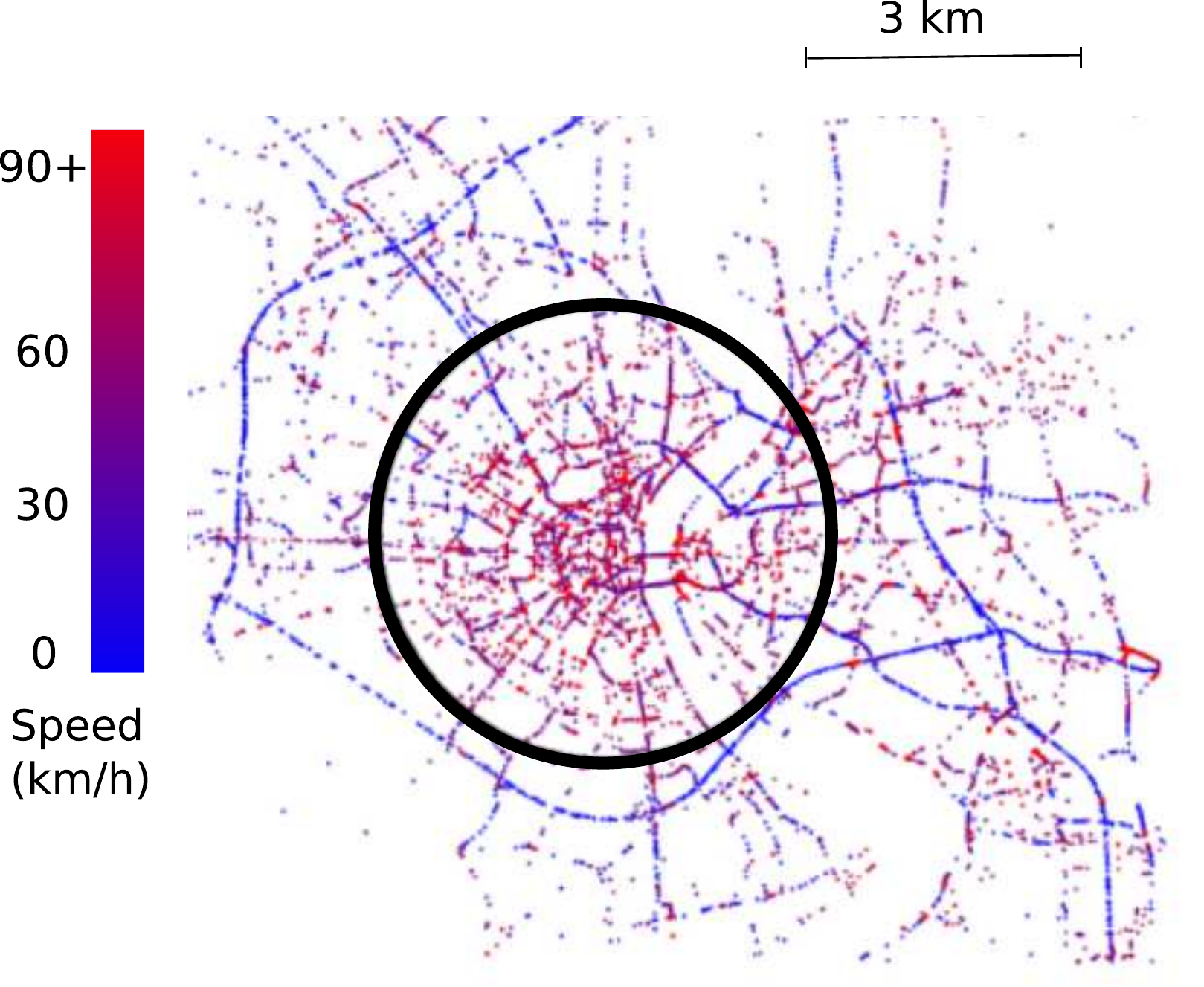}
        \caption{Cologne heat-map - adapted \cite{naboulsi2013}.}
        \label{fig:cologne-heatmap}
    \end{subfigure} 
     \caption{Cologne urban area.}
     \label{fig:cologne-topology-heatmap}
\end{figure}

In this simulation, we choose only a central submap of the city of Cologne.
The submap displays a higher incidence of traffic jams, as illustrated in Figure~\ref{fig:cologne-topology-heatmap}. 
Also, we consider that only vehicles send packages.
The residual distance objective is to compute the stability of the network~\cite{wahab2013vanet}.
We conduct all experiments thirty-three times on different traffic conditions with a confidence interval of 95\%. 
Table~\ref{table:simulation-parameters-settings} summarizes the simulation parameter settings.

\begingroup
\renewcommand\arraystretch{1.5}
\begin{table}[!htb]
\centering
\caption{Simulation parameters settings.}
\scalebox{0.9}{
\begin{tabular}{@{}ll@{}}
\toprule
\textbf{Parameter} & \textbf{Value} \\
\midrule
\textbf{Beacon transmission frequency} & 1Hz \\
\textbf{Total number of vehicles} & 13716 \\
\textbf{Transmission range} & 287m\\
\textbf{Density of vehicles} & [40–150] vehicles/$km^{2}$\\
\textbf{Bandwidth} & 10MHz \\
\textbf{Speed (average)}~\cite{Wahab2013} & 60.27 mph (96.9 km/h) \\
\textbf{Residual distance}~\cite{Wahab2013} & 63\% \\ 
\textbf{Channel} & 178 (5.89 GHz) \\
\textbf{MAC} & layerIEEE 802.11p PHY\\
\textbf{$\alpha$} & 0.5\\
\bottomrule
\end{tabular}}
\label{table:simulation-parameters-settings}
\end{table}
\endgroup

\begin{figure}[tb]
    \centering
    \begin{subfigure}[]{\tfig\textwidth}
        \includegraphics[width=1\textwidth]{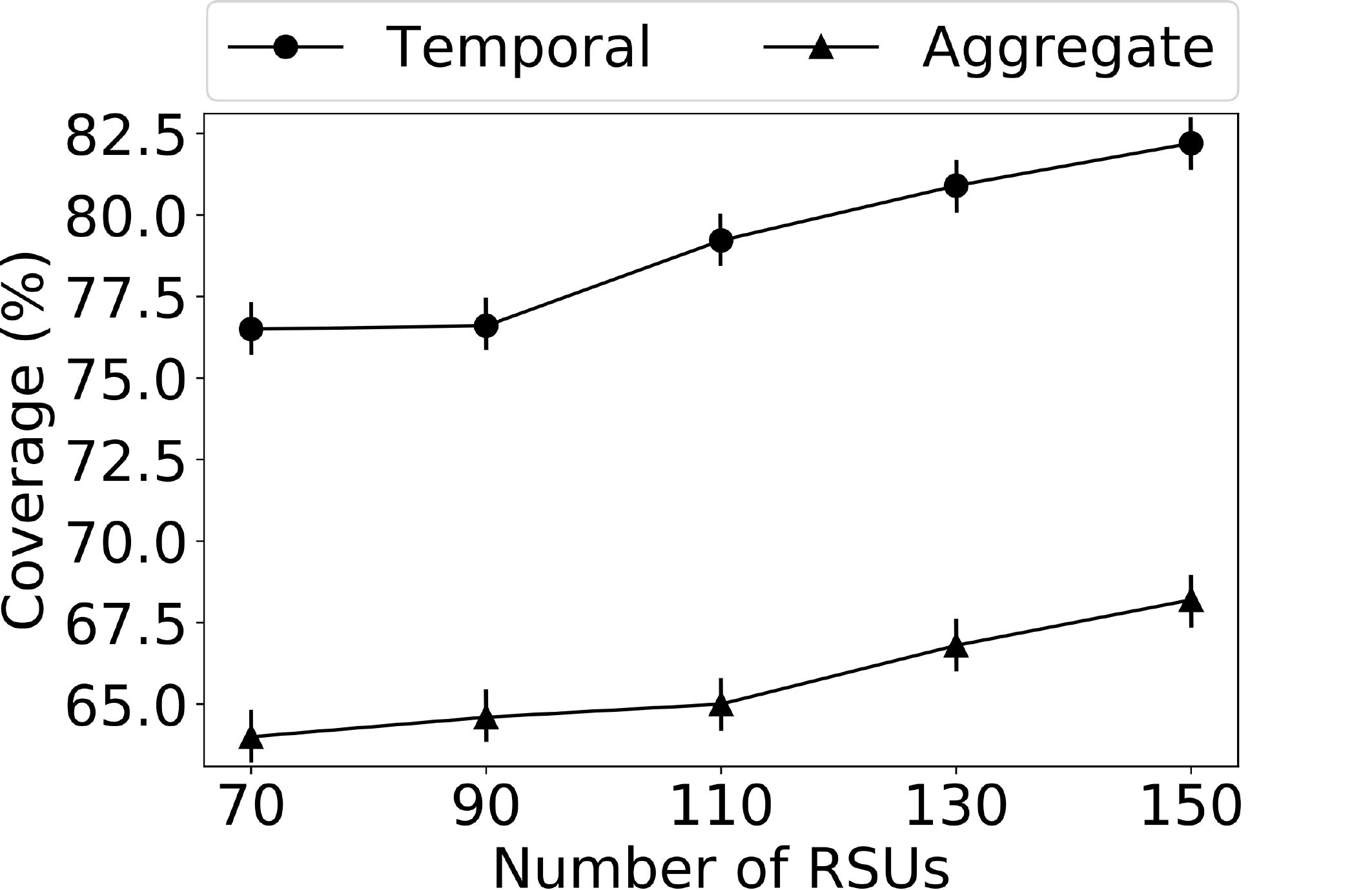}
        \caption{Coverage percentage.}
        \label{fig:coverage_rsu_simu}
    \end{subfigure}
    \begin{subfigure}[]{\tfig\textwidth}
    \includegraphics[width=1\textwidth]{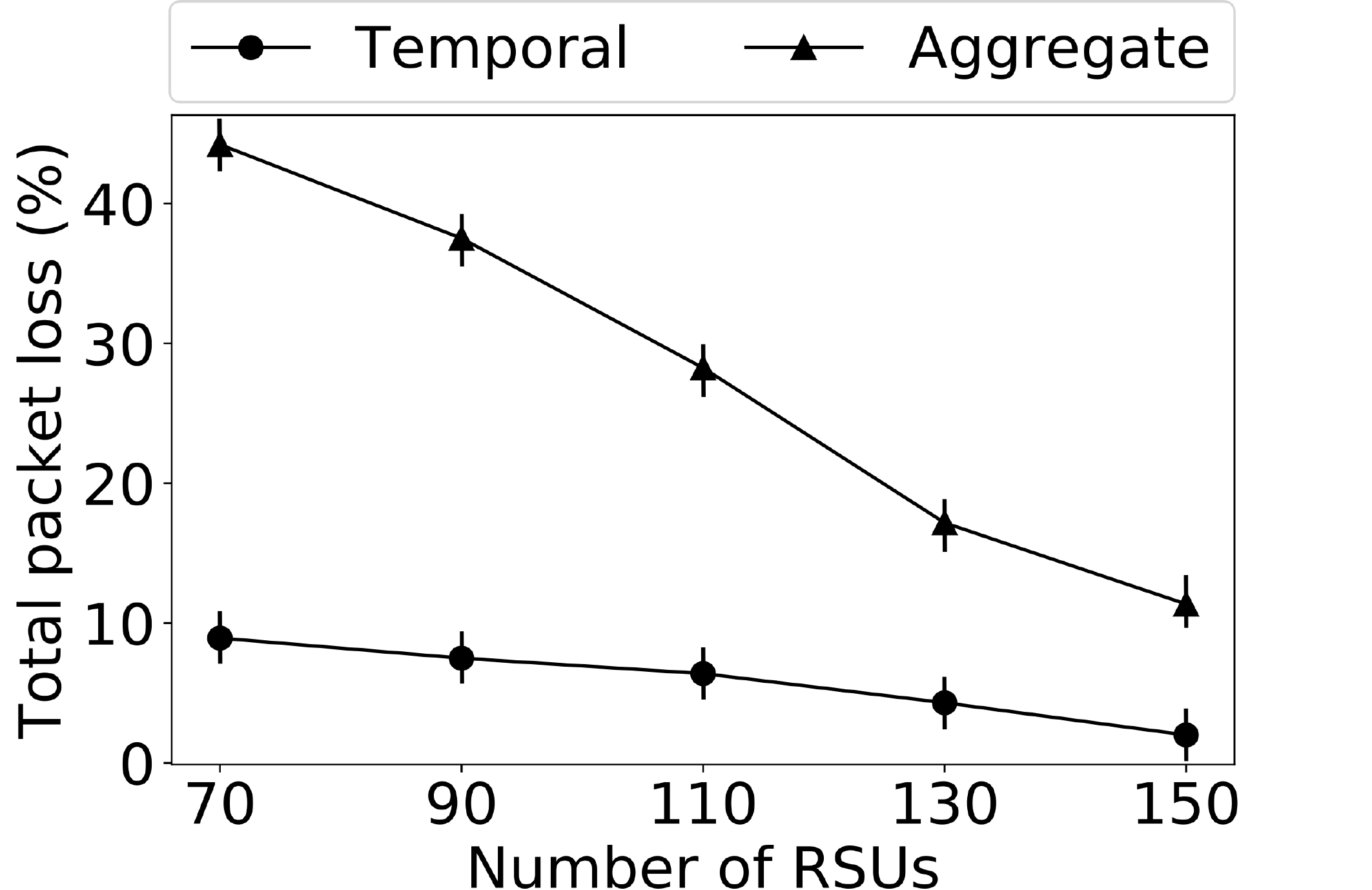}
        \caption{Total package loss.}
        \label{fig:total_package_loss_rsu_simu}
    \end{subfigure} 
     \caption{Coverage and Total package loss.}
     \label{fig:coverage_loss_simu}
\end{figure}

Figure~\ref{fig:coverage_loss_simu} presents the average vehicle coverage and total package loss with different $k$ values for both models.
As we can see in Figure~\ref{fig:coverage_rsu_simu}, for any RSU number, the coverage was higher when we use the temporal model.
For example, when $k=110$, the coverage is about $\approx 65\%$ with the aggregated model and $\approx 77\%$ with the temporal one using the same greed algorithm. 
For all cases, the greedy algorithm, in the aggregated model, selects the sets of RSUs, not considering the network dynamics and timing of temporal relationships.
In contrast, in the temporal model, the greedy algorithm selects the sets of RSUs, considering paths that are non-instantaneous between vehicle and RSU. 
They respect the network dynamics and timing of temporal relationships.

Figure~\ref{fig:total_package_loss_rsu_simu} shows the number of packet losses according to the RSU insertion rate. 
One factor that causes the rising of packet loss is improper RSUs
placement~\cite{Network-Connectivity-of-VANETs-in-Urban-Areas}.
For example, when $k=110$, the packet loss rate is $\approx 27\%$ with the aggregated and $\approx 8\%$ with the temporal model. 
In the aggregated model, the RSUs are concentrated in the central part of the scenario, covering a few regions and vehicles.
In contrast, in the temporal model, the RSUs are dispersed in the scenario, covering different regions and more vehicles than the aggregated model.
Therefore, the temporal model has a higher road coverage ratio and packet delivery than the aggregated model.

%% Alla - new post review section
\subsubsection{Temporal measures based solution}

Note that previous greedy heuristics provide a good approximation of the optimal solution but require global knowledge of the road topology and the vehicles that cross the intersections during the observation period.
We also compare the aggregated model and measures against the temporal ones using the genetic algorithm proposed by Moura et al.~\cite{moura2018evolutionary}.
They present a genetic algorithm strategy for the deployment of roadside units in VANETs and show that using a simple genetic algorithm with a betweenness centrality preprocessing could find a suitable solution when compared with other strategies.
The preprocessing uses the aggregated betweenness centrality.

We use the temporal model and measures in the preprocessing step of the genetic algorithm proposed by Moura et al..
The objective is to find the best intersection to install an RSU. 
The algorithm considers the intersections with the highest betweenness centrality (aggregated vs. temporal) to install an RSU.
The original algorithm~\cite{moura2018evolutionary} considers the aggregated Betweenness measure $B(v)$.
In this work, we also evaluate the degree $\kappa(v)$ and closeness $C(v)$ (aggregated vs. temporal). 

We consider the aggregated, and temporal graphs of the interval from $ 7 am $-$ 8 am $ (rush hour), in the Cologne scenario, with the 300 RSUs and contact time is $\unit[30]{seconds}$.
The temporal graph represents intervals of $\unit[30]{seconds}$, totaling ten-time graphs.
The results with the temporal model and measures were always superior to the aggregated ones.
\begin{itemize}
    \item $\kappa(v)$ presents coverage of 41.6\% with the aggregated model vs. 59.2\% with the use of temporal one. This behavior occurs because, different of aggregated degree, the temporal one choose the RSUs based a degrees balance and consistent during the time observed.
    \item $C(v)$ presents coverage of 39.8\% with the aggregated model vs. 79.6\% with the use of temporal one. This behavior occurs because the temporal one chooses the RSUs based on an actual vehicle’s proximity.  The aggregated model could identify nonexistent proximity. For example, a node is present only in the first moment, but new proximity interactions will happen with future nodes present in the same area.
    \item $B(v)$ presents coverage of 61.4\% with the aggregated model vs. 91.3\% with the use of temporal one. This difference occurs because of the same reasons given previously.
\end{itemize}
In summary, the approach with betweenness had the best result, more than 90\% of the coverage area. 
In this way, we could see the differences between the models and how they influence this decision.

%% Alla - Substitui pela texto 
\begin{comment}
Figure~\ref{fig:coverageRSU_with_measures} presents the percentage coverage area for aggregated model ($G_{10}^a$), temporal model ($G_{[0,10]}^t$) and aggregated applied in each intervals ($G_1,\ldots,G_{10}$). 
When we use the temporal model, the results are always superior to the aggregated one and almost better than aggregated applied in each interval. 
The approach with betweenness had the best result, more than 80\% of the coverage area.
\begin{figure}[!htb]
    \centering
    \begin{subfigure}[]{\tfig\textwidth}
        \includegraphics[width=\textwidth]{RSU/coverage_V2/coberturaRSU-dg-V2.pdf}
        \caption{Degree based solution}
        \label{fig:coverageRSU-degree}
    \end{subfigure} % \hspace{5mm} 
    \begin{subfigure}[]{\tfig\textwidth}
        \includegraphics[width=\textwidth]{RSU/coverage_V2/coberturaRSU-cls-V2.pdf}
        \caption{Closeness based solution}
        \label{fig:coverageRSU-cls}
    \end{subfigure} % \hspace{5mm} 
    \begin{subfigure}[]{\tfig\textwidth}
        \includegraphics[width=\textwidth]{RSU/coverage_V2/coberturaRSU-btw-V2.pdf}
        \caption{Betweenness based solution}
        \label{fig:coverageRSU-btw}
    \end{subfigure}
    \caption[RSU's coverage area with measures based solution.]{RSU's coverage area with measures based solution.}
    \label{fig:coverageRSU_with_measures}
\end{figure}
\end{comment}

\section{Conclusion and future work}\label{sec:conclusion}
In this work, we provide an essential analysis regarding the impact of modeling the VANETs as an aggregated and temporal graph.
To do so, we use centrality measures for temporal networks and discuss how the results can be affected by the topology modeling approach.

In the first analysis, we observe the variations of the network topology.
We realized that the aggregated modeling does not identify topological temporal information.
In the second analysis, we quantify the aggregated model's impact compared to the temporal one using the Kolmogorov-Smirnov test and Hellinger distance. 
We found a $p$-value $ < \num{2.2e-16}$ for all cases in Kolmogorov-Smirnov test, so we rejected the null hypotheses: $H_0:p_{G}(.) = p_{G^{[0,T]}}(.)$.
Unlike KS-test, Hellinger distance can better distinguish the impact of the aggregated model in the presented scenarios. 
We presumed that the shortest paths in the temporal graph and an aggregated graph are similar because the smallest value for the $\mathcal{H}^2_{B}$ was $0.0905$ in the Cologne scenario.
Finally, we analyzed our modeling through a scatter plot and showed how strongly the temporal measures are numerically related to each other.

%%%%%%%%%%%%
%% Alla - inclui novos comentários nas conclusões
To quantify the impact of temporal modeling in a dynamic and simulated environment, we consider the RSU allocation problem in Cologne's city's central submap.
First, we compare our strategy with a greedy algorithm to choose the RSUs' positions. 
The results show that for all $k$ values, the temporal model is always superior to the aggregated one using the same placement strategy.
For instance, in a scenario with 70 RSUs, we have 77\% and 65\% of coverage in the temporal and aggregated model, respectively. 
Second, we compare aggregated modeling against the temporal ones as features in a genetic algorithm based on complex network measures.
The approach with temporal betweenness had the better result with 90\% of the coverage area against 61\% of aggregated one applied to the same scenario.

An essential contribution of this work is to use temporal measures providing better solutions in a vehicular network environment. 
To the best of our knowledge, we are the only ones that use these measures in VANETs scenarios.
As future work,  we want to study more relationships in the vehicle environment, such as vehicle connectivity, traffic optimization, location of points of dissemination to maximize vehicle coverage, and analyze vehicle environments with V2X communication.
Moreover, we intend to use temporal centrality measures to identify the most viable anchor zones to spread floating contents in VANETs.

%% Alla - Essa parte é muito importante, se não tiver limite de páginas deixar.
\section*{Software availability}
The information about the modeling package software are the follows:
 \begin{itemize}
     \item Name of software: TC-VANETs.
     \item Developer contact: Fillipe dos Santos Silva \url{fillipe.silva@students.ic.unicamp.br}. 
     \item Address: Cidade Universit\'aria Zeferino Vaz, s/n - Bar\~ao Geraldo. CEP: 13083-970. Campinas - SP - Brazil. 
     \item Phone number: +55 (19) 3521-7000.
     \item Year first available: 2019.
     \item Hardware suggested: Processor \unit[2.6]{GHz} Intel Core i7, with \unit[16]{GB} \unit[1600]{MHz} DDR3 and HD \unit[1]{TB}. However, we performed the simulations under a computer model SGI Rackable Standard-Depth Servers, with 56 cores in 2 sockets of \unit[2.00]{GHz} Intel(R) Xeon(R) CPU E5-2660 v4, \unit[1352]{MHz} \unit[256]{GB} DDR3 Memory (32 $\times$ \unit[8]{GB}), and Linux operating system.
     \item Software required: All project is running over Ubuntu 16.04 (recommended) or 18.04.
     \item Availability: Available on Github 
 \footnote{\url{https://github.com/fillipesansilva/VANETs-topology-modeling-based-on-temporal-complex-networks}}
     \item Cost: All free tools.
 \end{itemize}

We implemented the project in Statistical R\footnote{\url{https://www.r-project.org}} and Python 2\footnote{\url{https://www.python.org/}}.

\section*{Acknowledgments}
The authors acknowledge support from the Brazilian research agency CNPq, and INCT of the Future Internet for Smart Cities (CNPq 465446/2014-0 and CAPES 88887.136422/2017-00).

%Padronizar os campos das referências.
\bibliographystyle{elsarticle-num}
\bibliography{main}

\end{document}